\documentclass[aps,prc,twocolumn,superscriptaddress,showpacs,nofootinbib]{revtex4-1}
\usepackage{graphicx,amsmath,amssymb,bm,color}

\newcommand{\mev}{\,\mathrm{MeV}}

\newcommand{\fm}{\,\mathrm{fm}}

\newcommand{\km}{\,\mathrm{km}}

\newcommand{\beq}{\begin{equation}}
\newcommand{\eeq}{\end{equation}}

\definecolor{dartmouth}{rgb}{0.05,0.5,0.06}

\allowdisplaybreaks

\begin{document}

\title{Spectrum of shear modes in the neutron-star crust: \\
Estimating the nuclear-physics uncertainties}
\author{I.\ Tews}
\email[E-mail:~]{itews@uw.edu}
\affiliation{Institute for Nuclear Theory,
University of Washington, Seattle, WA 98195-1550}

\begin{abstract}
I construct a model of the inner crust of neutron stars using 
interactions from chiral effective field theory (EFT) in order to 
calculate its equation of state (EOS), shear properties, and the 
spectrum of crustal shear modes. I systematically study 
uncertainties associated with the nuclear physics input, the crust 
composition, and neutron entrainment, and estimate their impact 
on crustal shear properties and the shear-mode spectrum. I find 
that the uncertainties originate mainly in two sources: The 
neutron-matter EOS and neutron entrainment. I compare the 
spectrum of crustal shear modes to observed frequencies of 
quasi-periodic oscillations in the afterglow of giant $\gamma$-ray bursts 
and find that all of these frequencies could be described within 
uncertainties, which are, however, at present too sizable to infer 
neutron-star properties from observations.
\end{abstract}

\maketitle

\section{Introduction}\label{sec:introduction}

Neutron stars are remarkable objects: With masses up to 
2~M$_{\odot}$~\cite{Demorest2010, Antoniadis2013} and typical radii 
of the order of 12 km~\cite{Lattimer:2014, Hebeler:2013, Watts:2016uzu},
densities inside neutron stars are higher than densities accessible in 
experiments on earth. This makes neutron stars excellent laboratories 
for physical theories under extreme conditions. 

A large part of the available observational data on neutron stars is linked 
to the physics of neutron-star crusts, which can be divided into the outer 
and the inner crust. The outer crust consists of a lattice of neutron-rich 
nuclei emerged in a sea of electrons. Deeper in the neutron star, with 
increasing density and neutron chemical potential, the nuclei become more 
and more neutron-rich. At densities of $\rho \sim 4\times 10^{11} 
\text{g/cm}^3$,  the neutron chemical potential becomes positive and 
neutrons begin to drip out of the nuclei. This is where the inner crust 
begins. In addition to free neutrons, inhomogeneous phases of nuclear 
matter, the so-called nuclear pasta phases, may appear; see, e.g., 
Ref.~\cite{Schneider:2013dwa}. At the crust-core transition density, 
which is roughly half of the nuclear saturation density 
$\rho_{0}\sim 2.7\times 10^{14} \text{g/cm}^3 \sim 0.16 \fm^{-3}$, 
the nuclei will dissolve and a phase of uniform nuclear matter in $\beta$
equilibrium will begin.

Understanding crustal properties is key to describe various neutron-star
observations~\cite{Chamel:2008ca}. In this paper, I focus on shear 
properties of the neutron-star crusts: The shear modulus $\mu$, shear 
velocities $v_S$, and the frequency spectrum of crustal shear modes. 
Crustal shear modes are of particular interest for the description of
quasiperiodic oscillations (QPOs) in the afterglow of giant $\gamma$-ray 
bursts in magnetars~\cite{Israel:2005, Strohmayer:2005, Watts:2006,
Strohmayer:2006, Hambaryan:2010}. The shear modulus describes how 
the neutron-star crust elastically deforms under shear stress, i.e., it 
describes the stiffness of the crust lattice under shear deformations. 
These deformations lead to the formation of shear oscillations, which 
travel through the crust with the shear velocity $v_S$ and have a 
frequency that depends on $v_S$ and the crust parameters. Giant 
flares trigger starquakes that cause crustal shear deformations 
and lead to shear oscillations in the crust~\cite{Duncan:1998my}. 
These shear oscillations can in principle modulate the surface emission 
and then be observed as QPOs. However, QPOs are not simply crustal 
oscillation modes because the global magnetic field couples the neutron 
star's crust and core and leads to the formation of global oscillation 
modes~\cite{Levin:2006ck, Gabler:2010rp}. The global magnetic field, 
thus, plays an important role for the correct description of QPO 
oscillation spectra.

Because the restoring force in the crustal lattice is the Coulomb 
interaction, the shear modulus depends on the charge number $Z$ of 
the lattice ions and their density $n_i$. While for the outer crust these 
are well understood, the composition and structure of the inner crust are 
not well constrained. Furthermore, additional effects in the inner crust 
are thought to be crucial for the correct description of crustal shear modes, 
like neutron superfluidity~\cite{Andersson:2008, Samuelsson:2009xz, 
Passamonti:2011mc, Sotani:2013jya},
entrainment of neutrons with the crust lattice~\cite{Chamel:2012zn}, or 
the appearance of pasta phases~\cite{Gearheart:2011qt, Sotani:2011nn,
Passamonti:2016jfo}, but these effects are not completely understood.

This ignorance of crustal properties will also reflect in the crustal shear 
spectra. So far, no oscillation model, neither crustal nor global, is able to 
describe all observed QPO frequencies, see, e.g., Ref.~\cite{SteinerWatts2009}. 
On the other hand, none of these models include systematic uncertainties. 
In this paper, I estimate the effects of nuclear-physics uncertainties on
the spectrum of crustal shear modes. These uncertainties may be sizable 
and originate from various sources, e.g., the inner-crust EOS, the crust 
structure and composition, or neutron entrainment.

This paper is structured as follows: In Sec.~\ref{sec:crustEOS} I will 
determine models for the inner-crust equation of state (EOS) within the 
Wigner-Seitz
approximation, based on realistic interactions with systematic theoretical
uncertainties. I use these EOSs in Sec.~\ref{sec:shearspeed} to determine 
the shear modulus and the shear velocities of the neutron-star inner crust. 
Finally, in Secs.~\ref{sec:QPOs} and~\ref{sec:n1}, I calculate the frequencies
of the fundamental crustal shear oscillations as well as of the first radial 
overtone with nuclear-physics uncertainties with the goal of identifying the 
largest sources of uncertainty. I summarize and give an outlook in 
Sec.~\ref{sec:outlook}.

\section{Inner-crust equation of state}\label{sec:crustEOS}

I use a Gibbs construction within the Wigner-Seitz approximation to 
determine an inner-crust EOS consistent with realistic models of the 
nuclear interactions. For the outer crust, I will use the model by Baym, 
Pethick, and Sutherland (BPS) of Ref.~\cite{BPS} with the sequence of 
nuclei as calculated in Ref.~\cite{Ruester:2005}. Since the shear spectrum 
is largely insensitive to the outer-crust EOS and the exact neutron drip 
density~\cite{SteinerWatts2009}, this choice will not affect my main 
results. 

To describe the nuclear interactions in the inner crust, I will use two 
different parametrizations. First, I use an empirical parametrization
suggested in Ref.~\cite{Hebeler:2013} fit to realistic interactions from 
chiral effective field theory (EFT)~\cite{Epelbaum:2009a, Entem:2011}. 
Chiral EFT is a systematically improveable framework to describe low-energy
hadronic interactions, and is directly connected to the symmetries of 
quantum chromodynamics. It naturally includes both two-body and 
many-body forces, which are key for the correct description of nuclei and 
nuclear matter, see Ref.~\cite{Hebeler:review2015} and references therein. 
Due to its systematics, chiral EFT allows for systematic uncertainty 
estimates, which enables one to investigate the effects of uncertainties in 
the nuclear interactions on the crustal shear spectrum. Chiral EFT has been 
very successfully used in calculations of nuclear matter~\cite{Hebeler:2010a,
nucmatt, Holt2:2012, Tews:2013, CCnucmatt, Drischler:2013, Hagen:2014,
Roggero:2014, Wlazlowski:2014jna, Lynn2014, Carbone:2014, Lynn:2015, 
Drischler:2015} and nuclei~\cite{SM, CC1, CC1b, Cipollone:2013zma,
Hergert:2013b, Holt:2011fj, Holt:2010yb, Ca, Hergert:2014iaa, 
Soma:2013xha, nuclattice2, Hagen:2015yea, Ruiz:2016gne}, and was also 
used to study neutron-star properties~\cite{Hebeler:2013,Kruger:2013kua}. 

Second, I will explore the simpler parametrization suggested in 
Ref.~\cite{Gandolfi:2012}, where the parameters were fit in 
Ref.~\cite{Lattimer:2015nhk} both to phenomenological interactions
of the Argonne + Urbana/Illinois type as well as to a set of chiral EFT 
interactions.

\subsection{Inner crust from chiral EFT interactions}\label{sec:InnerCrust}

In the Wigner-Seitz approximation, one considers a spherical Wigner-Seitz
cell of pure neutron matter, phase I, with radius $R_W$ and volume 
$V_{W}$. In its center, I assume a nucleus in form of a spherical drop 
of asymmetric nuclear matter, phase II, with radius $R_{0}$, volume
$V_{0}$, and proton number $Z$. The proton density is constant inside
the nucleus, $n_p^C(\textbf{r})=e \,n_p \,\Theta(r-R_0)=e Z/V_{0} \,
\Theta(r-R_0)$. A relativistic electron gas is equally distributed in the 
Wigner-Seitz cell. The whole system is considered at $T=0$. 

For the two phases to be stable, the following two conditions need to be 
fulfilled: First, the pressure $P(n,x)=n^2 \partial(E/A)/\partial n$ in both 
phases has to be equal, $P(n,x)=P^I (n_n,0)=P^{II} (n,x)$, with the neutron
number density in phase I, $n_n$, and the baryon number density $n$ and 
proton fraction $x=n_p/n$ in phase II. Second, the neutron chemical 
potential, $\mu_{n}=\mu_p+\mu_{el}$, has to be equal in both phases. 
Since the Wigner-Seitz cell is immersed in a uniform electron gas, the 
equilibrium conditions are not affected by the presence of electrons, and 
one can include their contributions later.

The neutron and proton chemical potentials $\mu_n$ and $\mu_p$ are 
given by
\begin{align}
\mu_p(n,x)&=n \frac{\partial \left(\frac{E}{A}\right)}{\partial n}
+ \frac{\partial \left(\frac{E}{A}\right)}{\partial x}(1-x)+\frac{E}{A}+m_p\,, \\
\mu_n(n,x)&=n \frac{\partial \left(\frac{E}{A}\right)}{\partial n}
- \frac{\partial \left(\frac{E}{A}\right)}{\partial x}x+\frac{E}{A}+m_n\,,
\end{align}
with the neutron and proton masses $m_n$ and $m_p$, respectively. The
pressure and chemical potential can be derived easily from the energy 
per particle $E/A$, given in phase I by
\begin{align}
\frac{E}{A}(n_n,0)=\frac{E^{\text{nuc}}}{A}(n_n,0)\,,\label{eq:EN}
\end{align}
and in phase II by
\begin{align}
\frac{E}{A}(n,x)=\frac{E^{\text{nuc}}}{A}(n,x)+\frac{E^C}{A}(n,x)+\frac{E^S}{A}(n,x)\,.\label{eq:ENII}
\end{align}
The term $E^{\text{nuc}}$ takes into account the kinetic energy 
of the nucleons as well as the nuclear interaction energy, $E^C$ is the 
Coulomb energy, and $E^S$ the surface energy of the nuclear drop. 

I first make use of the parametrization of Ref.~\cite{Hebeler:2013}, 
given by
\begin{align} \label{eq:parametrization}
\frac{E^{\text{nuc}}}{A}(n,x)&= T_0  \left(\frac35\left(x^{\frac53}
+(1-x)^{\frac53} \right)\left(\frac{2n}{n_0}\right)^{\frac23}\right.\\
&\quad - \left[(2\alpha - 4\alpha_L) x (1 - x) +  \alpha_L \right] \frac{n}{n_0} \nonumber \\
&\quad \left. + \left[(2\eta - 4\eta_L) x (1 - x) +  \eta_L \right] \left(\frac{n}{n_0}\right)^{\gamma} \right) \nonumber\,,
\end{align}
where $n_0$ is the nuclear saturation density, $n_0=0.16 \, \text{fm}^{-3}$,
$T_0=(3 \pi^2 n_0/2)^{2/3}\hbar^2/(2 m_N)$ is the Fermi energy of 
symmetric nuclear matter at saturation density, and $\gamma=4/3$. The 
parameters $\alpha, \alpha_L, \eta$, and $\eta_L$ are determined from fits 
to the empirical saturation point, $\frac{E^{\text{nuc}}}{A}(n_0,0.5)=-16 
\mev$, and $P^{\text{nuc}}(n_0,0.5)=0$, and to the neutron-matter 
results of Ref.~\cite{Hebeler:2010a} from chiral EFT. The full parameter 
range can be found in Ref.~\cite{Hebeler:2013}, and it allows one to quantify 
the uncertainties of modern nuclear interactions.

This parametrization leads to an incompressibility parameter $K=236 \mev$ 
and a skewness parameter $K'=-384 \mev$. The symmetry energy $S_v$ 
and its density dependence at saturation density, $L$, defined via 
\begin{align}
S_v(n)&=\left.\frac{1}{8} \frac{\partial^2}{\partial x^2}\frac{E}{A}\left(n,x\right)\right|_{x=1/2} \,,\\
L&=\left.3n_0\frac{\partial}{\partial n}S_V(n)\right|_{n_0}\,,
\end{align}
are in the ranges $S_v=(29.7-33.2) \mev$ and $L=(32.5-57) \mev$, respectively.
These ranges are in excellent agreement with experimental constraints (see, 
e.g., Ref.~\cite{Lattimer2012}), which rule out much larger values of L. 
Moreover, the parametrization of Eq.~\eqref{eq:parametrization} is in
remarkable agreement with explicit asymmetric-matter 
calculations~\cite{Hebeler:2013, Drischler:2013} and, thus, can be used 
with confidence to describe both neutron as well as asymmetric nuclear 
matter. 

\begin{figure}[t]
\centering
\includegraphics[trim= 0 0 0 0cm, clip=,width=0.45\textwidth]{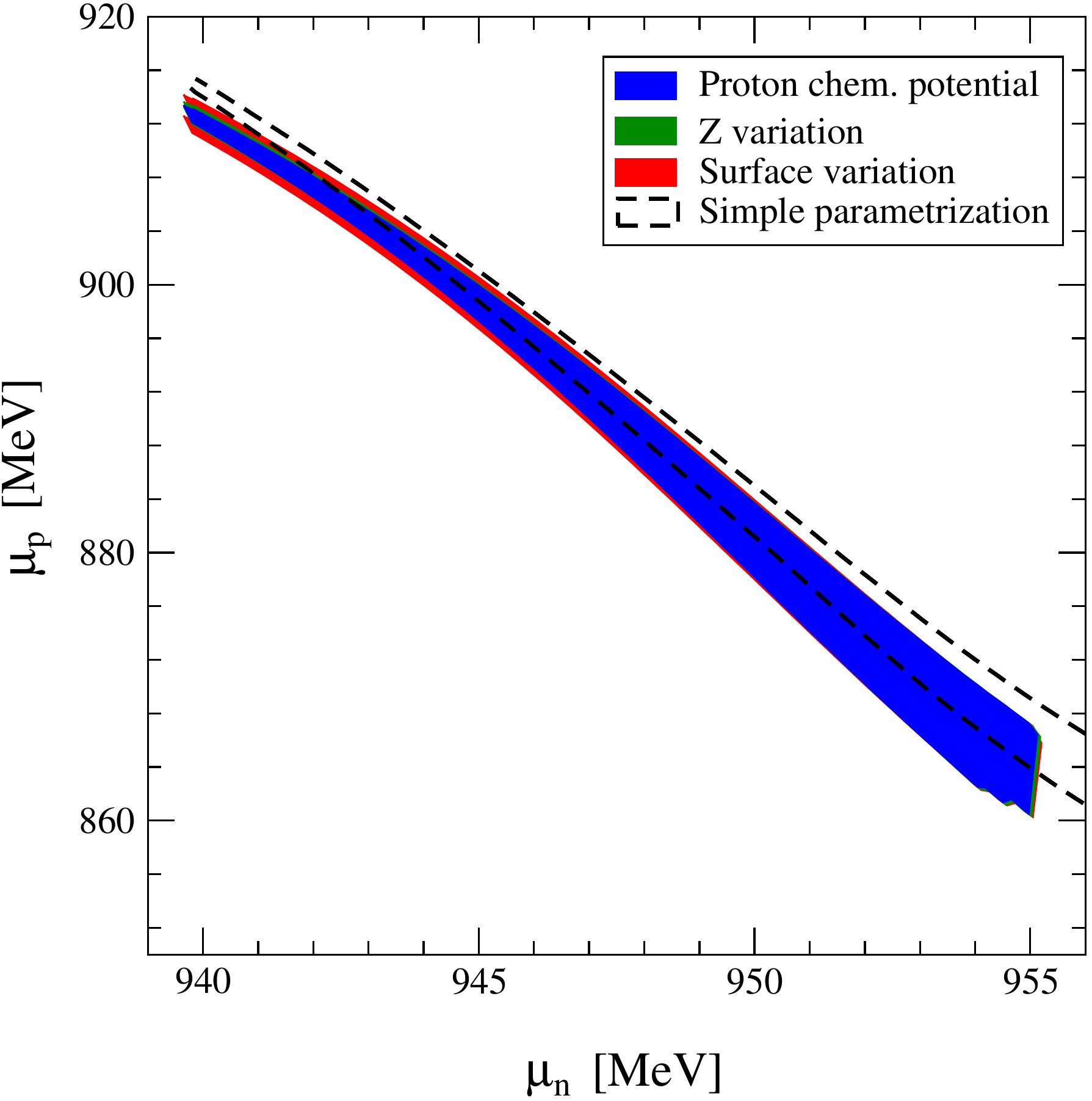}
\caption{Proton chemical potential as a function of the neutron chemical 
potential. The different bands show the uncertainties due to different 
sources: the variation of the surface parameters (red band), of Z (green 
band), and of the neutron-matter EOS (blue band). As one can see, the 
uncertainties are dominated by the blue band (which overlaps the red and 
green bands almost always completely). I also show the results
obtained by using a simpler parametrization for the nuclear interaction 
(dashed bands); see Sec.~\ref{sec:simpleEOS}. 
 \label{fig:munmup}}
\end{figure}

\begin{figure}[t]
\centering
\includegraphics[trim= 0 0 0 0cm, clip=,width=0.45\textwidth]{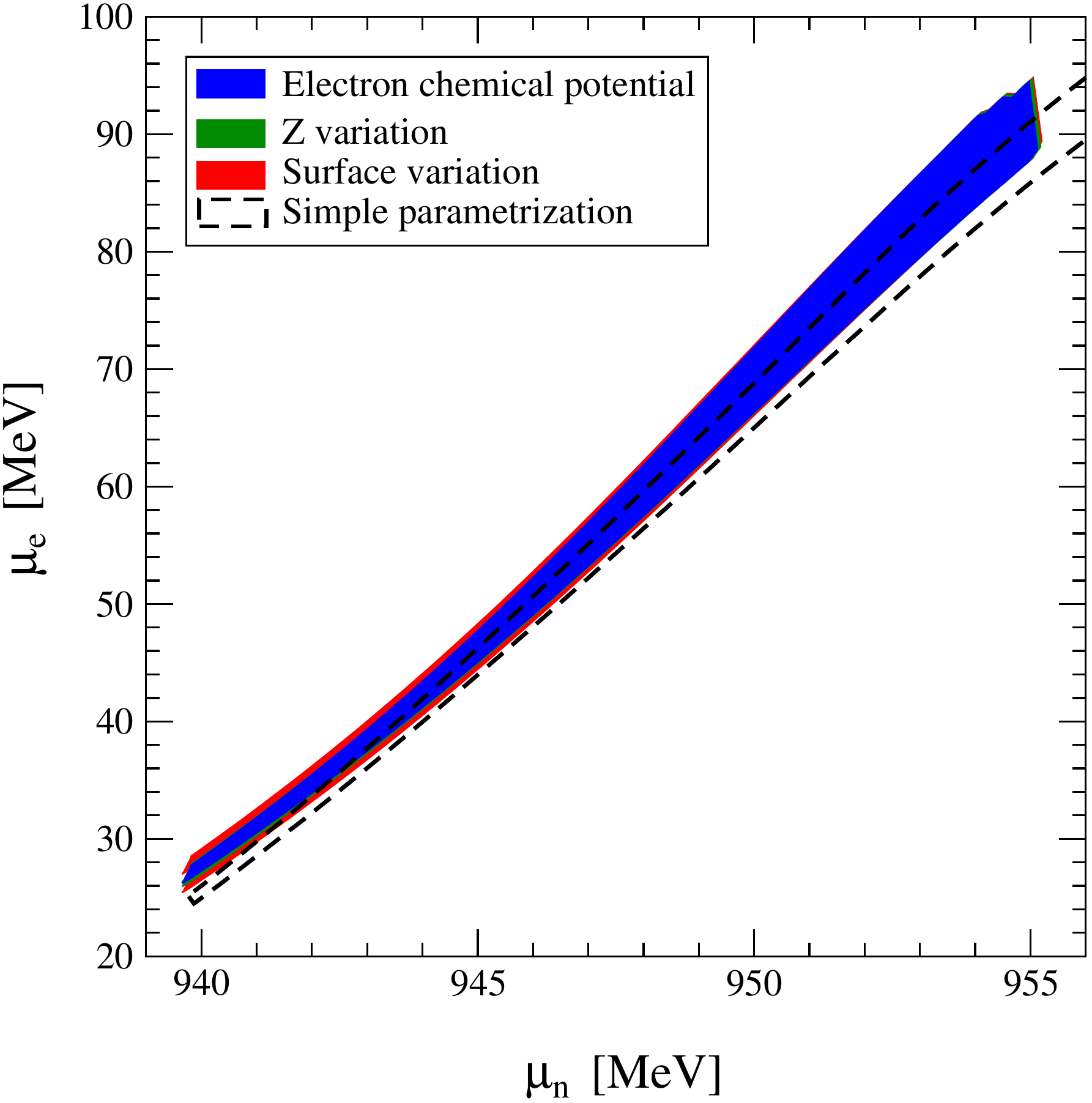}
\caption{Electron chemical potential as a function of the neutron chemical 
potential, where the bands are obtained as in Fig.~\ref{fig:munmup}. 
 \label{fig:munmue}}
\end{figure}

The Coulomb energy per particle of a nucleus with uniformly distributed
protons is given by
\begin{align}
\frac{E^C}{A}(n,x)&= \frac35 \alpha_{\text{FS}} \, \hbar c \left(\frac{4\pi}{3}\right)^{\frac13} Z^{\frac23} \,n^{\frac13} x^{\frac43}\,, \label{eq:coul}
\end{align}
with the fine structure constant $\alpha_{\text{FS}}$. This form is 
analogous to the Coulomb term in the phenomenological Bethe-Weizsaecker 
mass formula. 

The surface energy $E^S$ is the product of surface area and surface 
tension $\sigma(x)$,
\begin{align}
\frac{E^S}{A}(n,x)&= \frac{1}{A} \sigma(x) \cdot 4 \pi R_0^2= \sigma(x) \cdot \left(\frac{36 \pi x}{ Z n^2} \right)^{\frac13}\,.
\end{align}
I follow Ref.~\cite{Steiner:2007} and expand the surface tension in 
the neutron excess $\beta=(1-2x)$,
\begin{align}
\sigma(x)=\sigma_0(1-\sigma_{\beta} (1-2x)^2 + \cdots ) \,,\label{eq:surfII}
\end{align}
where $\sigma_{\beta}$ is the symmetry parameter of the surface tension. 
Because the surface tension measures the energy needed to support the 
surface against the lower density in the outer phase, it has to vanish for 
$x\to 0$ in the neutron-star inner crust (because in this case both phases
are identical). Then, one can modify the surface tension,
\begin{align}
\sigma(x)=\sigma_0\frac{16+b}{\frac{1}{x^3}+b+\frac{1}{(1-x)^3}}\,,
\label{eq:surfI}
\end{align}
where $b=96/\sigma_{\beta}-16$. To determine the surface parameters 
$\sigma_0$ and $\sigma_{\beta}$, I fit the binding energies of 
Eq.~\eqref{eq:ENII} with the surface tension as defined in Eq.~\eqref{eq:surfII} 
to the experimentally measured masses from Ref.~\cite{Audi:2014}. From
$\sigma_{\beta}$ I determine $b$, and use the surface tension of
Eq.~\eqref{eq:surfI} for the inner crust of the neutron star. 

I account for uncertainties in the modeling and the fitting procedure 
and vary both $\sigma_0$ and $b$ within a $10 \%$ uncertainty. This
uncertainty estimate is large enough to bring the binding energies of 
Eq.~\eqref{eq:ENII} into agreement with experimental values for all 
nuclei. 

In addition to the nucleonic contributions to the system, I consider a 
free relativistic electron gas with constant electron density $n_{el}$ in 
the Wigner-Seitz cell. The electron energy density is given 
by~\cite{Chamel:2008ca}
\begin{align}
\varepsilon_{el}(n_{el})&=\frac{m_{el}^4 c^8}{8 \pi^2 \hbar^3 c^3}
\left(x_r(2 x_r^2+1)(x_r^2+1)^{\frac12}\right. \\ \nonumber
&\quad \left.-\ln(x_r+(x_r^2+1)^{\frac12})\right)\,,
\end{align}
with $x_r= \hbar k_F / (m_{el} c) =(3 \pi^2 n_{el})^{1/3}\,\hbar c/m_{el}$ 
and the electron mass $m_{el}=0.511$ MeV. For the free electron gas, 
this leads to an electron chemical potential of $\mu_{el}^{\text{free}}=
\left(\hbar^2 c^2 (3 \pi^2 n_{el})^{2/3} + m_{el}^2c^4  \right)^{\frac12}$ 
and an electron pressure, $P_{el}(n_{el})=n_{el} \partial \varepsilon_{el}/ 
\partial n_{el}-\varepsilon_{el}$.

The contribution of the electrons to the Coulomb energy is given by two
terms. The first one is the lattice energy~\cite{Chamel:2008ca},
\begin{align}
\varepsilon^{C}_L(n_{el})=-\frac{9}{10} \alpha_{FS} \hbar c \left(\frac{4\pi}{3}\right)^{\frac13} Z^{\frac23} n_{el}^{\frac43} \,,
\end{align}
and describes the electron Coulomb energy assuming pointlike nuclei. 
The second term corrects for the finite size of the nuclei, and is given 
by~\cite{Chamel:2008ca}
\begin{align}
\varepsilon_L^{\text{C,corr}}(n_{el})=-\frac13 \varepsilon^{C}_L(n_{el}) w^{\frac23}\,,
\end{align}
where $w=\frac{V_{0}}{V_W}$ is the volume fraction of the nucleus in 
the Wigner-Seitz cell. One can rewrite the electron Coulomb contribution 
in terms of $n$ and $x$ and include it in $E^C$ in Eq.\eqref{eq:coul}, which 
leads to
\begin{align}
\frac{E^C}{N}(n,x)&= \frac35 \alpha_{FS} \hbar c \left(\frac{4\pi}{3}\right)^{\frac13} Z^{\frac23} \,n^{\frac13} x^{\frac43}\\ \nonumber
& \quad \times \left(1-\frac32 w^{\frac13}+ \frac12 w \right)\,.
\end{align}

\begin{figure}[t]
\centering
\includegraphics[trim= 0 0 0 0cm, clip=,width=0.45\textwidth]{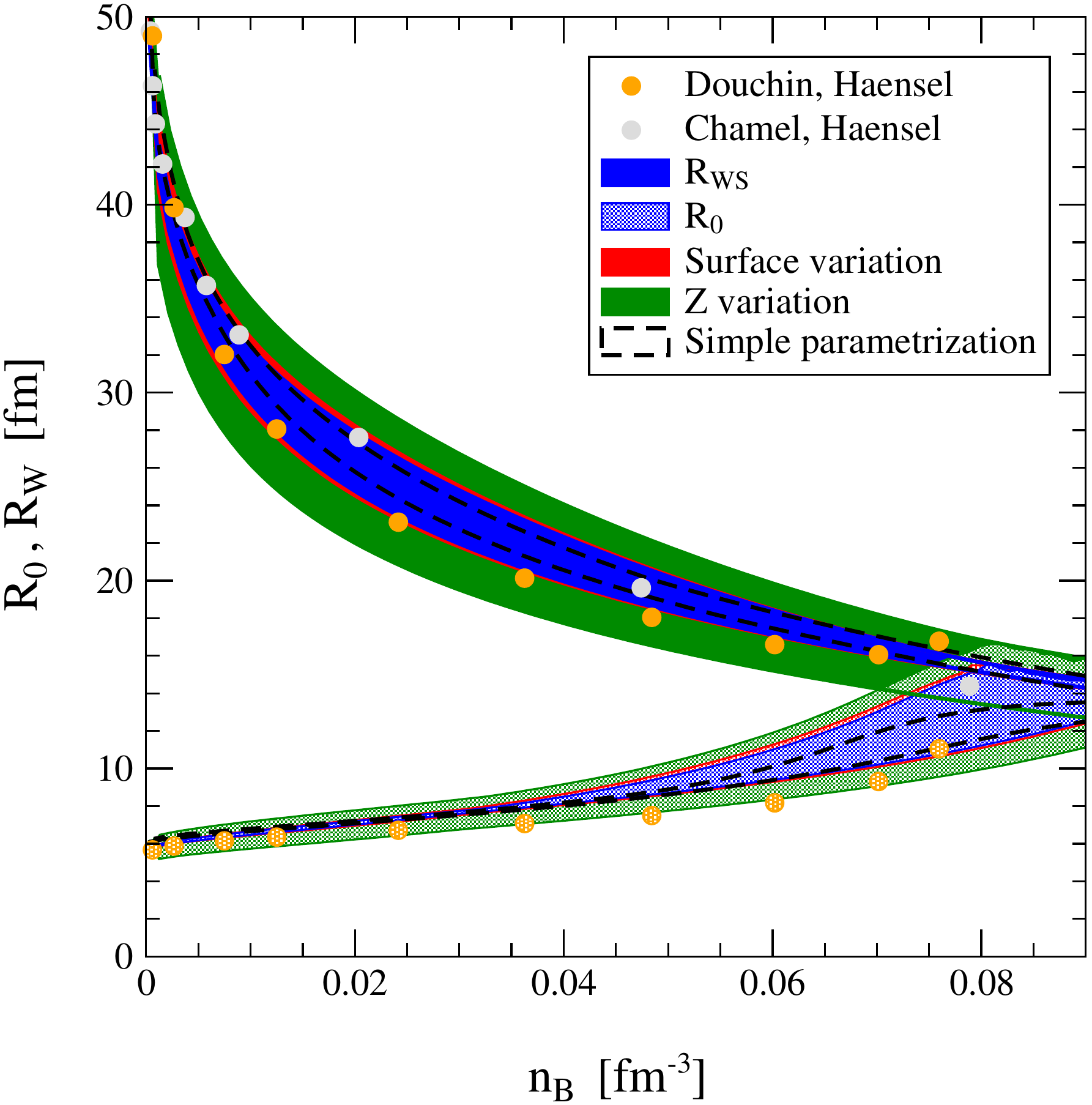}
\caption{Radii of the nucleus $R_0$ (dotted, lower bands) and of the 
Wigner-Seitz cell $R_W$ (filled, upper bands) as a function of baryon density. 
The bands are determined as in Fig.~\ref{fig:munmup}. I also show results 
by Douchin and Haensel~\cite{Douchin:2001sv} (orange points) and 
$R_{W}$ from the calculation by Chamel and Haensel~\cite{Chamel:2008ca} 
(grey points).
 \label{fig:radii}}
\end{figure}

\begin{figure}[t]
\centering
\includegraphics[trim= 0 0 0 0cm, clip=,width=0.45\textwidth]{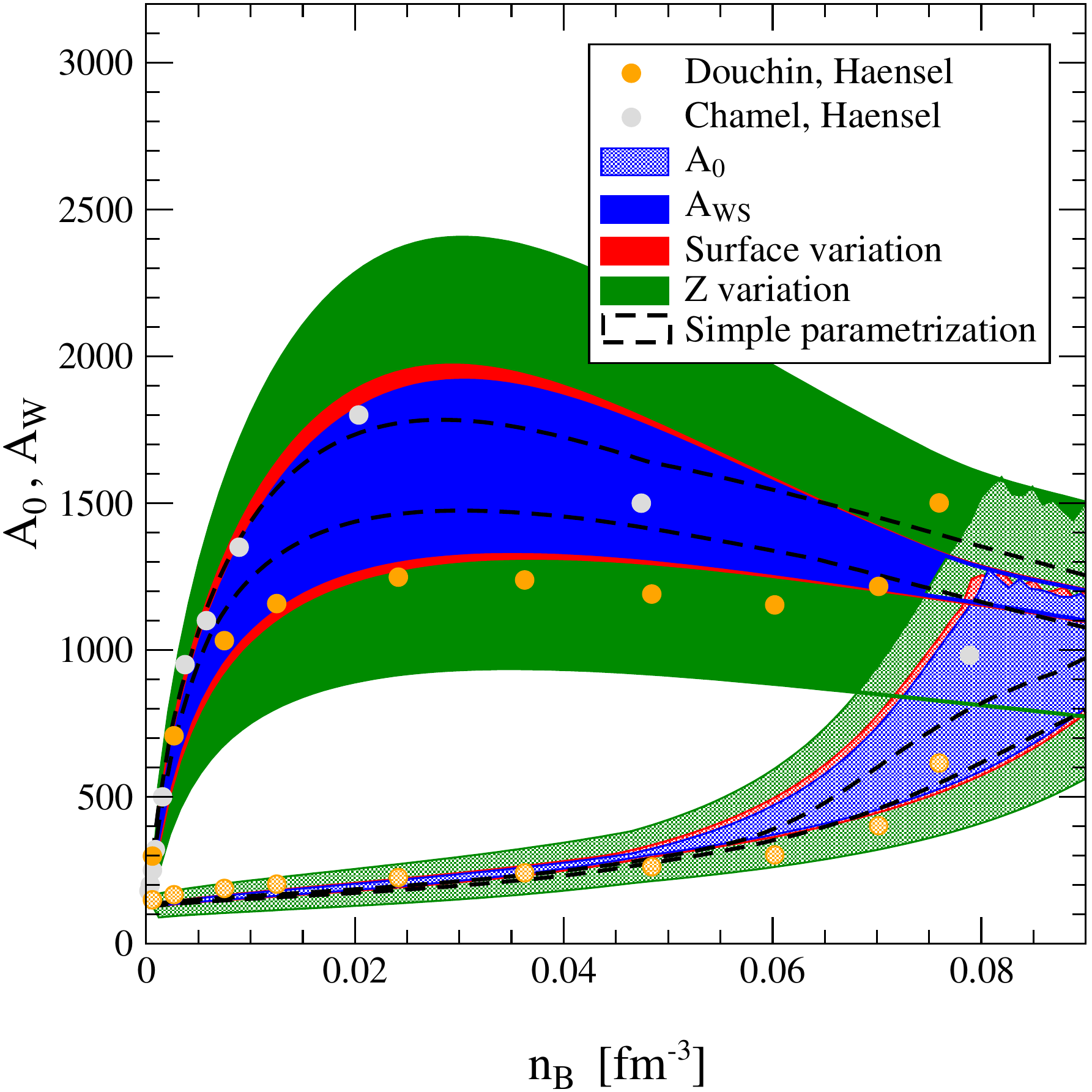}
\caption{Nucleon numbers of the nucleus, $A_0$, (dotted, lower bands) and 
of the Wigner-Seitz cell, $A_W$, (filled, upper bands) as a function of baryon 
density. The bands are determined as in Fig.~\ref{fig:munmup}. I also show
results by Douchin and Haensel~\cite{Douchin:2001sv} (orange points) and
$A_{W}$ from the calculation by Chamel and Haensel~\cite{Chamel:2008ca} 
(grey points).
\label{fig:AAWS}}
\end{figure}

To obtain the radii of the nucleus and the Wigner-Seitz cell, one first needs
to determine the proton number $Z$ of the nuclei. $Z$ can be obtained 
by minimizing the energy inside the Wigner Seitz cell, which is complicated
as shell effects have to be considered. As a simplification, I assume that 
nuclei in the neutron-star crust favor configurations with closed proton 
shells~\cite{Negele:1971vb} and choose $Z=40$. To account for uncertainties
associated with this assumption, I vary $Z$ in a range $Z=28-50$, which
includes the two neighboring shell closures, and overlaps with various 
determinations; see Ref.~\cite{Chamel:2008ca}.

For each neutron density $n_n$, the equilibrium conditions constrain 
density and proton fraction in the nuclear phase. By varying $n_n$, and 
thus, fixing the pressure and the neutron chemical potential, one can 
determine the composition and density of the nucleus as well as the 
electron chemical potential $\mu_{el}=\mu_n-\mu_p$ and, thus, the 
electron density. In Figs.~\ref{fig:munmup} and \ref{fig:munmue}, I 
show the proton and electron chemical potentials $\mu_{p}$ and 
$\mu_{e}$ as functions of $\mu_{n}$. In addition to the nuclear-physics
uncertainties from the neutron-matter EOS (blue band), I show the
uncertainties due to the $Z$ variation (green band) and due to variation 
of the surface parameters (red band). 

The main uncertainty in the chemical potentials stems from the variation 
of the neutron-matter EOS. This is to be expected, because pressure and
chemical potentials are mostly set by the bulk EOS. The $Z$ variation 
has only a small effect and can be neglected even at low densities (low 
neutron chemical potential). $Z$ primarily impacts the Coulomb energy, 
which is the smallest contribution to the energy at all densities and rapidly
decreases with increasing density. The uncertainty in the surface parameters 
is considerable only at low densities. At higher densities, $\sigma(x)$ (and,
thus, the surface energy) decreases and its uncertainty can be neglected. 

With known $Z$ and proton density inside the nucleus, one can determine
the volume of the nucleus, $V_0=Z/n_p=Z/(n x)$. To determine the size 
of the Wigner-Seitz cell, $V_W$, I enforce charge neutrality, $Q=0=
V_0 n_p -V_W n_{el}$, and obtain $V_W=Z/n_{el}$. The radii $R_0$ and 
$R_W$ are plotted in Fig.~\ref{fig:radii}, and the nucleon numbers $A_0$ 
and $A_W$ inside the nucleus and the Wigner-Seitz cell, respectively, are 
shown in Fig.~\ref{fig:AAWS}, with similar uncertainty bands as before. It 
is clear that the choice of the proton number $Z$ has the largest impact 
on these numbers. I compare my calculation to the results of 
Ref.~\cite{Douchin:2001sv} and with $A_W$ and $R_W$ obtained in 
Ref.~\cite{Negele:1971vb} and find very good agreement.

Although I calculate all crustal properties up to the density at which the 
Gibbs construction breaks down, I need to determine the crust-core 
transition density, $n_{cc}$, to calculate the shear spectrum. The crust-core 
transition density is the density at which the Gibbs free energy of the 
crustal lattice becomes larger than that of uniform nuclear matter in 
$\beta$ equilibrium. For my crust model I find that the transition density
ranges from $0.074$ to $0.090 \fm^{-3}$.  The exact value of the crust-core
transition density in principle influences the QPO frequencies calculated in 
Secs.~\ref{sec:QPOs} and \ref{sec:n1}, because it impacts the crust thickness 
$\Delta R$. While the fundamental mode frequencies, which mainly scale 
with the neutron-star radius, are almost independent of the exact transition
density, the overtones depend on the its exact value~\cite{SteinerWatts2009,
Deibel:2013sia}. However, since I later also vary the crust thickness in a 
sizable range, any additional uncertainty in the transition density will be 
accounted for.

Having determined all parameters of the Wigner-Seitz cell, one can model 
the crust as a Coulomb lattice of Wigner-Seitz cells, with the density of
nuclei, $n_i$, given by $n_i=(4/3 \pi R_W^3)^{-1}$. At higher densities 
in the inner crust, though, matter may form nuclear pasta phases, and the
Wigner-Seitz approximation will fail. I do not consider these phases in 
this work, which may be an additional source of uncertainty, see, e.g., 
Ref.~\cite{Gearheart:2011qt, Sotani:2011nn, Passamonti:2016jfo}. While the appearance 
of nuclear pasta phases will not affect the equilibrium conditions (pressure 
and neutron chemical potential at a certain $n_n$ will be unchanged), it 
may affect the composition, energy density, and shear properties at higher
densities in the inner crust. The elastic properties of these phases are still
unknown~\cite{Passamonti:2016jfo} and their impact needs to be studied 
in future work.

\subsection{Inner crust from a simple parametrization}\label{sec:simpleEOS}

I also explore a simpler parametrization for the energy per particle and
start from the four-parameter parametrization for pure neutron matter, 
given by~\cite{Gandolfi:2012}
\begin{align}
\frac{E}{N}(n,0)=a\left(\frac{n}{n_0} \right)^{\alpha}+b\left(\frac{n}{n_0} \right)^{\beta}\,,
\end{align}
where the parameters $a$ and $\alpha$ describe the low-density equation 
of state, while the parameters $b$ and $\beta$ determine the higher-density 
part of the EOS. I consider the parameter range given in 
Ref.~\cite{Lattimer:2015nhk}, which includes fits of these four parameters 
both to phenomenological interactions of the Argonne and Urbana/Illinois 
type as well as to selected chiral neutron matter EOSs. This simple 
parametrization leads to a similar neutron-matter EOS compared to the 
empirical parametrization used before.

To describe the energy in the nuclear cluster, I expand the energy 
per particle for asymmetric matter around symmetric nuclear matter,
\begin{align}
\frac{E}{A}(n,x)&=\frac{E}{A}(n,x=0.5)+S_v(n)(1-2x)^2+ ...\\ \nonumber
&=-16\mev+\frac12 \frac{\partial^2 \frac{E}{A}}{\partial n^2} (n-n_0)^2 \\ \nonumber
&\quad+\frac16 \frac{\partial^3 \frac{E}{A}}{\partial n^3} (n-n_0)^3+ ...\\ \nonumber
&\quad+S_v(n)(1-2x)^2+ ...\,,
\end{align}
where I have expanded around the saturation point. I choose the 
incompressibility and skewness parameters
\begin{align}
K&=9 n_0^2 \frac{\partial^2 \frac{E}{A}}{\partial n^2}=236\mev\,, \\ \nonumber
K'&=27 n_0^3 \frac{\partial^3 \frac{E}{A}}{\partial n^3}=-384\mev\,,
\end{align}
consistent with the values for the previous parametrization, and approximate 
the symmetry energy as $S_v(n)=\frac{E}{N}(n,0)-\frac{E}{A}(n,x=0.5)$. 
For the Coulomb energy, I use the same form as before. For the simple 
model, to further reduce the number of parameters, I choose the surface 
energy to be $E^S/A=2 E^C/A$~\cite{Ravenhall:1983}.

The results for this simpler model are shown in each plot as a black-dashed 
band. For the proton chemical potentials in Fig.~\ref{fig:munmup} (electron
chemical potentials in Fig.~\ref{fig:munmue}), the results for the simple 
parametrization lie at the upper (lower) boundary of the results for the
empirical parametrization. At low densities, the two parametrizations start 
to diverge, due to the simplified modeling of the surface energy. At higher
densities, the simple model leads to higher (lower) proton (electron) 
chemical potentials due to the inclusion of fits to the Argonne+Urbana 
interactions, which give higher neutron-matter pressure than chiral models.

These effects are reflected in the radii and nucleon numbers in 
Figs.~\ref{fig:AAWS} and~\ref{fig:radii}. A higher proton chemical 
potential for the same neutron chemical potential leads to a higher 
proton density in the nucleus. Thus, for proton chemical potentials at
the upper boundary of the empirical parametrization (and a constant 
$Z$) one would expect radii and nucleon numbers at the lower boundary.
Similarly, for slightly lower electron chemical potentials one would expect 
slightly larger radii and nucleon numbers inside the Wigner-Seitz cell.

\subsection{Equation of State}

\begin{figure}[t]
\centering
\includegraphics[trim= 0 0 0 0cm, clip=,width=0.45\textwidth]{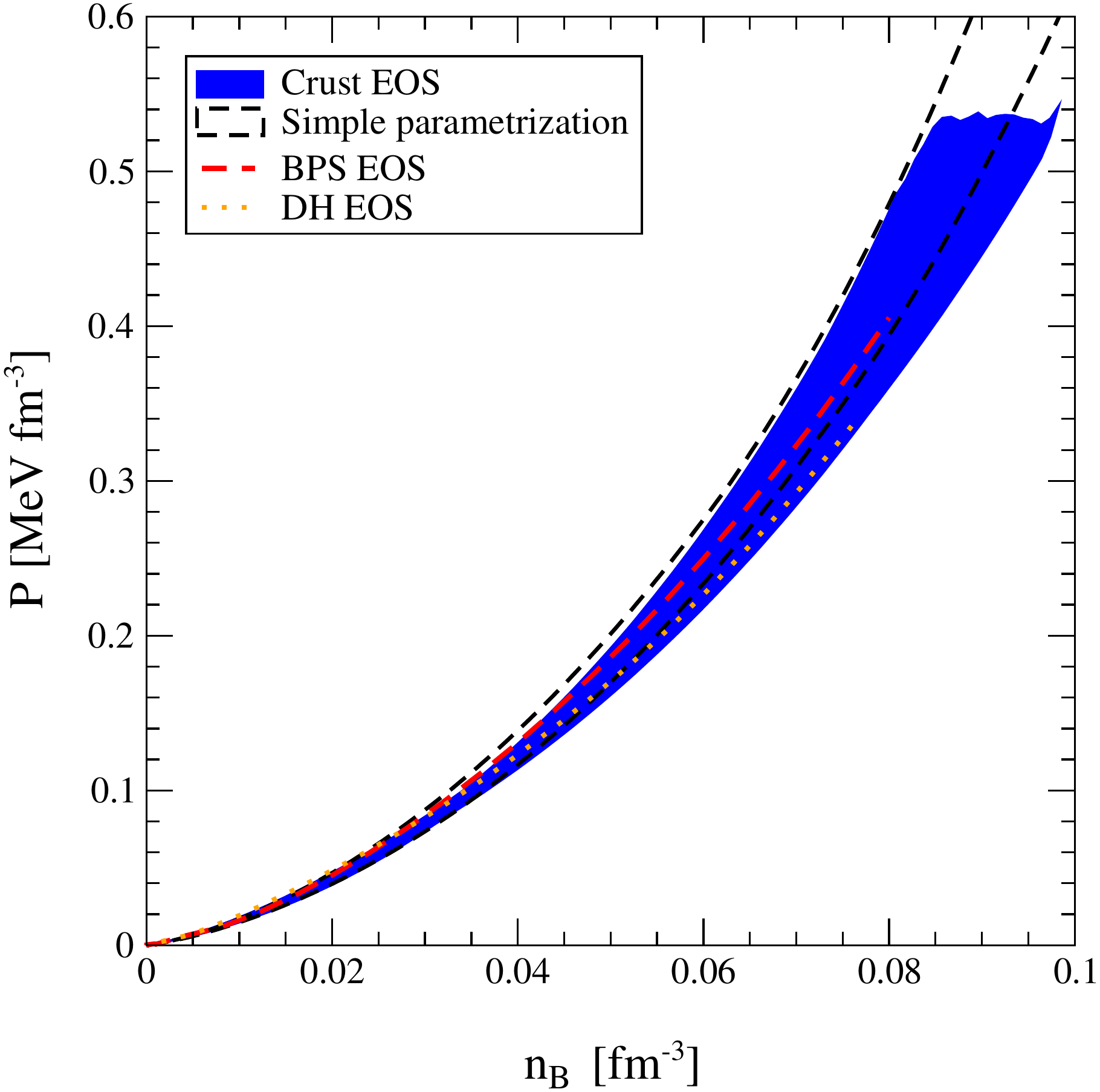}
\caption{Inner-crust EOS (pressure as a function of baryon density) for 
the inner-crust model of this work. The uncertainty band for $P(n_B)$ 
solely originates in the uncertainty of the neutron-matter EOS. I also 
show the BPS EOS~\cite{BPS} and the EOS by Douchin and 
Haensel~\cite{Douchin:2001sv}.
\label{fig:crustEoS}}
\end{figure} 

As described in Sec.~\ref{sec:InnerCrust}, for every input neutron density 
$n_n$ in the neutron-matter phase one can determine the energy density 
in the Wigner-Seitz cell,
\begin{align}
\varepsilon_W &=w\cdot n\frac{E}{A}(n,x) \\ \nonumber
&\quad + (1-w) \cdot n_n \frac{E}{A}(n_n,0)+ \varepsilon_{el}(n_{el})\,,
\end{align}
as well as the baryon density $n_B=w\cdot n + (1-w)\cdot n_n$, and the
total pressure $P=P_{I}+P_{el}=P_{II}+P_{el}$. One, thus, can obtain the 
EOS of the inner crust, $P=P(\varepsilon)$ or $P=P(n_B)$, with theoretical
uncertainties. The inner-crust EOS is plotted in Fig.~\ref{fig:crustEoS} and 
agrees well with the BPS EOSs~\cite{BPS} and the EOS by Douchin and 
Haensel~\cite{Douchin:2001sv}.

The only source of uncertainty for the inner-crust EOS is the neutron-matter 
EOS. This is intuitive because for a given input density in the neutron-matter
phase, $n_n$, the pressure is set independently of $Z$ or the surface 
parameters. At the same time, the baryon density is determined primarily 
by the neutron-matter density: Variation of the surface and Coulomb 
energies, by variation of $Z$ and the surface parameters, is only relevant 
at low densities, where the neutron-matter phase dominates the volume of 
the Wigner-Seitz cell. The main effect of the $Z$ variation is a change of 
the volumes of the Wigner-Seitz cell and the nucleus by the same factor and,
thus, these volume changes do not affect the energy or baryon density. 

When calculating the equation of state with the simple parametrization, 
one finds very good agreement with the empirical parametrization, see 
Fig.~\ref{fig:crustEoS}. This is expected, as both models describe the 
neutron-matter EOS similarly as well as the empirical saturation point. 

\begin{figure}[t]
\centering
\includegraphics[trim= 0 0 0 0cm, clip=,width=0.47\textwidth]{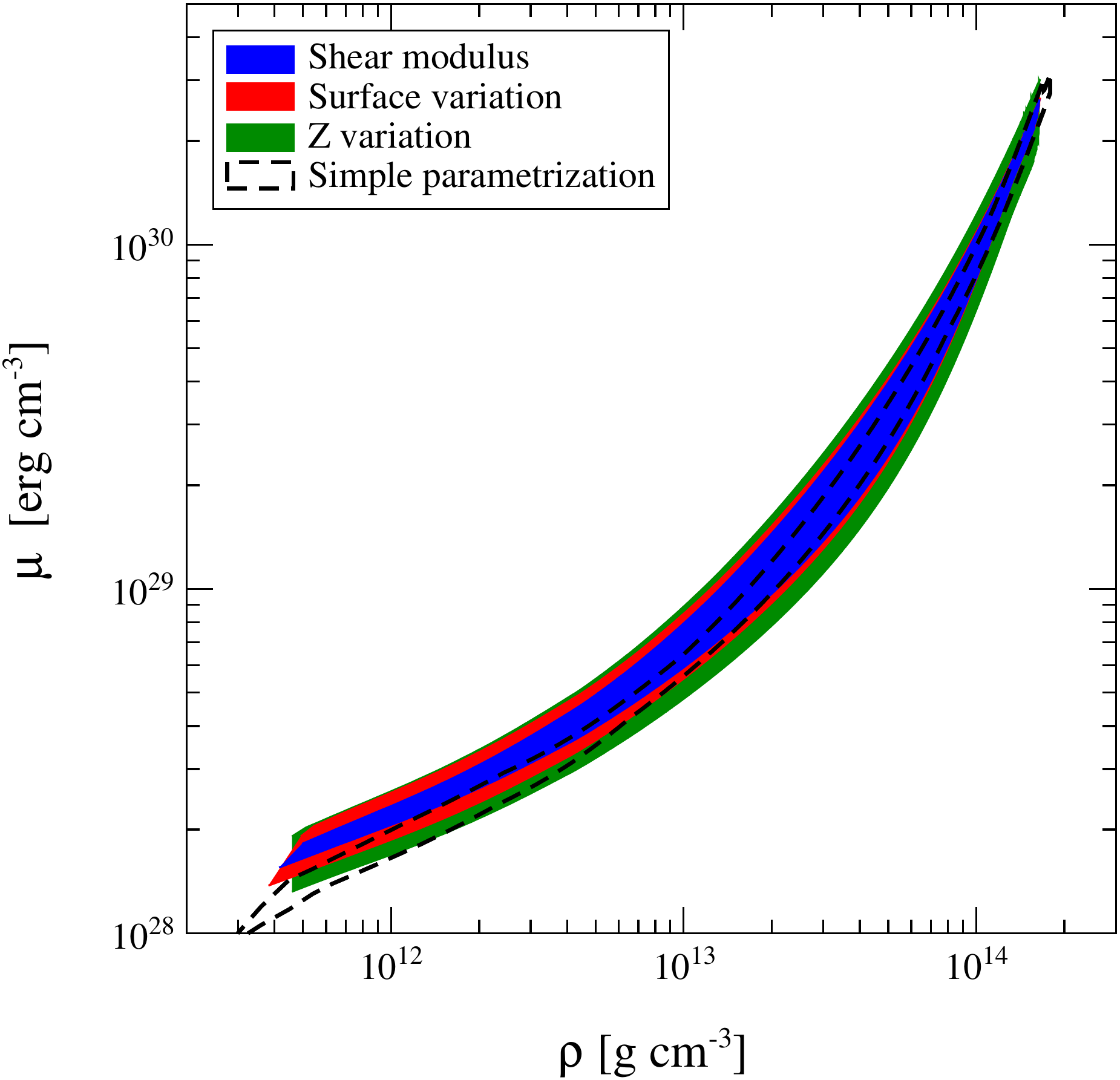}
\caption{Shear modulus as a function of mass density, where the uncertainty
bands are obtained as in Fig.~\ref{fig:munmup}.
 \label{fig:ShearMod}}
\end{figure} 

\section{Shear modulus and shear velocities}\label{sec:shearspeed}

The shear modulus for a body-centered cubic Coulomb lattice in the 
neutron-star crust is given by Ref.~\cite{Strohmayer:1991}
\begin{align}
\mu=\frac{0.1194}{1+0.595 (\Gamma_0/\Gamma)^2}\frac{n_i (Ze)^2}{a}\,.
\end{align}
Here, $n_i$ is the density of nuclei, and $a=(3/(4 \pi n_i))^{1/3}$. 
The parameter $\Gamma=(Ze)^2/a k_B T$ with temparature T and 
the Boltzmann constant $k_B$ is the ratio of Coulomb and thermal 
energies. The upper boundary of the crust is defined as $\Gamma_0=
173$~\cite{Farouki:1993}. Assuming $T=0$ and including electron 
screening effects~\cite{Kobyakov:2013}, $\mu$ is given by
\begin{align}
\mu=0.1194 \left(1-0.010 Z^{\frac23}\right) \frac{n_i (Ze)^2}{a}\,.
\label{eq:shearmod}
\end{align}
I use this form for the shear modulus throughout the whole crust and 
neglect possible effects of nuclear pasta phases on $\mu$, which could 
be sizable~\cite{Gearheart:2011qt, Sotani:2011nn, Passamonti:2016jfo}.

Using the properties of the Wigner-Seitz cell, $n_i=1/V_W$, and $a=
R_W$, I obtain the shear modulus in the neutron-star inner crust and
show the results in Fig.~\ref{fig:ShearMod}. Since the shear modulus is 
purely geometrical and only depends on $R_W$, the uncertainties in the 
EOS and in Z are dominant. Using the simple parametrization, I find 
very good agreement at higher densities. For lower densities, at the 
top of the inner crust, both parametrizations start to disagree, which 
is a direct consequence of the higher $R_W$; see Fig.~\ref{fig:radii}. 

The shear velocities in the crust follow from the shear modulus, $v_S= 
(\mu/\rho_c)^{\frac12}$, with $\rho_C$ being the dynamical mass 
density (which is the mass density of nucleons moving with the lattice). 
When neglecting the effects of neutron superfluidity, the dynamical mass 
density equals the total mass density, $\rho_C=
\rho$~\cite{SteinerWatts2009}. Neutron superfluidity, however, plays an
important role in neutron star modeling~\cite{Andersson:2008} because 
free superfluid neutrons typically do not add to the dynamical mass density.
They are unlocked from the movement of the lattice and do not affect 
the shear properties of the crust. This effect can reduce $\rho_C$ 
considerably compared to $\rho$~\cite{Andersson:2008}, leading to
larger shear velocities~\cite{Sotani:2013jya}. 

In reality, the unbound superfluid neutrons, however, still interact 
with the lattice due to Bragg scattering, which can effectively lock a 
considerable portion of them in the lattice (entrained neutrons).
Chamel~\cite{Chamel:2012zn} found that up to 90\% of unbound 
neutrons could be entrained with the lattice, and, thus, only a small 
fraction of neutrons would be effectively free (conduction neutrons). 
The density of entrained neutrons, $n_n^{ent}$, the density of all 
unbound neutrons, $n_n^{ub}$, and the density of conduction neutrons, 
$n_n^c$, are related,
\begin{align}
n_n^{ent}=n_n^{ub}-n_n^c=n_n^{ub}\left(1-\frac{n_n^c}{n_n^{ub}}\right)=n_n^{ub}\cdot R_e\,,
\end{align}
where I define $R_e$ as the fraction unbound neutrons that are 
entrained. I use the density-dependent values for entrainment from 
Ref.~\cite{Chamel:2012zn}, which were determined using Skyrme 
potentials. Because these values are model dependent and not 
consistently derived for the chiral interactions I use throughout this 
work, I will associate large uncertainties with them. I will vary 
$R_e$ starting from no entrainment at all, $n_n^c/n_n^{ub}=1, 
R_e=0$, up to full entrainment of all neutrons, $n_n^c/n_n^{ub}=0, 
R_e=1$.  The latter case is equivalent to setting $\rho_C=\rho$ as in 
Ref.~\cite{SteinerWatts2009}. This range of variation of $R_e$ is the 
same as in Ref.~\cite{Sotani:2013jya}. While it appears to be large, the values
for $R_e$ calculated in Ref.~\cite{Chamel:2012zn} themselves vary 
between $0.3$ and $0.9$ in the density range relevant for the 
neutron-star inner crust. Furthermore, Kobyakov and 
Pethick~\cite{Kobyakov:2013}
suggested a sizable correction factor of $\approx 0.4$ to these values.

The variation in $R_e$ leads to a variation of shear velocities, with the 
minimum obtained for $R_e=1$ and the maximum for $R_e=0$. Please 
note that the shear modulus defined in Eq.~\eqref{eq:shearmod} is purely
geometrical and, thus, does not depend on entrainment. 

In Fig.~\ref{fig:ShearSpeed}, I show the shear velocities for the values 
of $R_e$ from Ref.~\cite{Chamel:2012zn} (blue band). In addition to the 
uncertainty bands from $Z$ variation (green band) and variation of the 
surface parameters (red band), I show the variation with $R_e$ (light-red 
band) and present a combined uncertainty band (grey band). I find that 
the variation of the entrainment parameter has the major impact on the 
shear velocities: They vary within a factor of 2 in the density regime of
interest when varying $R_e=0-1$. The range of shear velocities agrees 
very well with the range found in Ref.~\cite{Sotani:2013jya} for $L=73$,
which is close to the maximal $L$ found for chiral interactions. Entrainment,
thus, will also have a large impact on the shear spectra, which I discuss in 
the next sections. 

For the simple parametrization and $R_e$ values from 
Ref.~\cite{Chamel:2012zn}, I find good agreement of both 
parametrizations, which was expected based on the shear moduli. I also
compare my results to the calculation of Ref.~\cite{SteinerWatts2009}
(turqoise band), which was obtained using $R_e=1$. The shear properties 
of the crust are sensitive to the symmetry energy, $S_v$, and its density
dependence parameter, $L$. In particular, the shear velocities are 
anticorrelated with the $L$ parameter. Because the chiral interactions have
smaller $L$ values compared to some of the Skyrme models used in 
Ref.~\cite{SteinerWatts2009}, the shear velocities for chiral interactions
lie at the upper boundary of the band in Ref.~\cite{SteinerWatts2009}. 
Considering this, both results are in very good agreement.

\section{Frequencies of the fundamental crustal shear mode}\label{sec:QPOs}

Magnetars may provide a possibility of obtaining neutron-star shear
properties by investigating giant $\gamma$-ray bursts. These bursts are 
triggered when the very strong magnetic fields of the magnetars decay
over time and create unstable field-line configurations. Since the field 
lines are pinned to the magnetar's crust, unstable configurations may 
exert stress on the crust, which, at a certain point, ruptures and allows 
the magnetic field to reconfigure~\cite{Lander:2014}. While the crust
rupturing produces neutron starquakes, the field-line reconfigurations 
create currents which dissipate and produce giant $\gamma$-ray 
bursts~\cite{Duncan:1998my}.

\begin{figure}[t]
\centering
\includegraphics[trim= 0 0 0 0cm, clip=,width=0.45\textwidth]{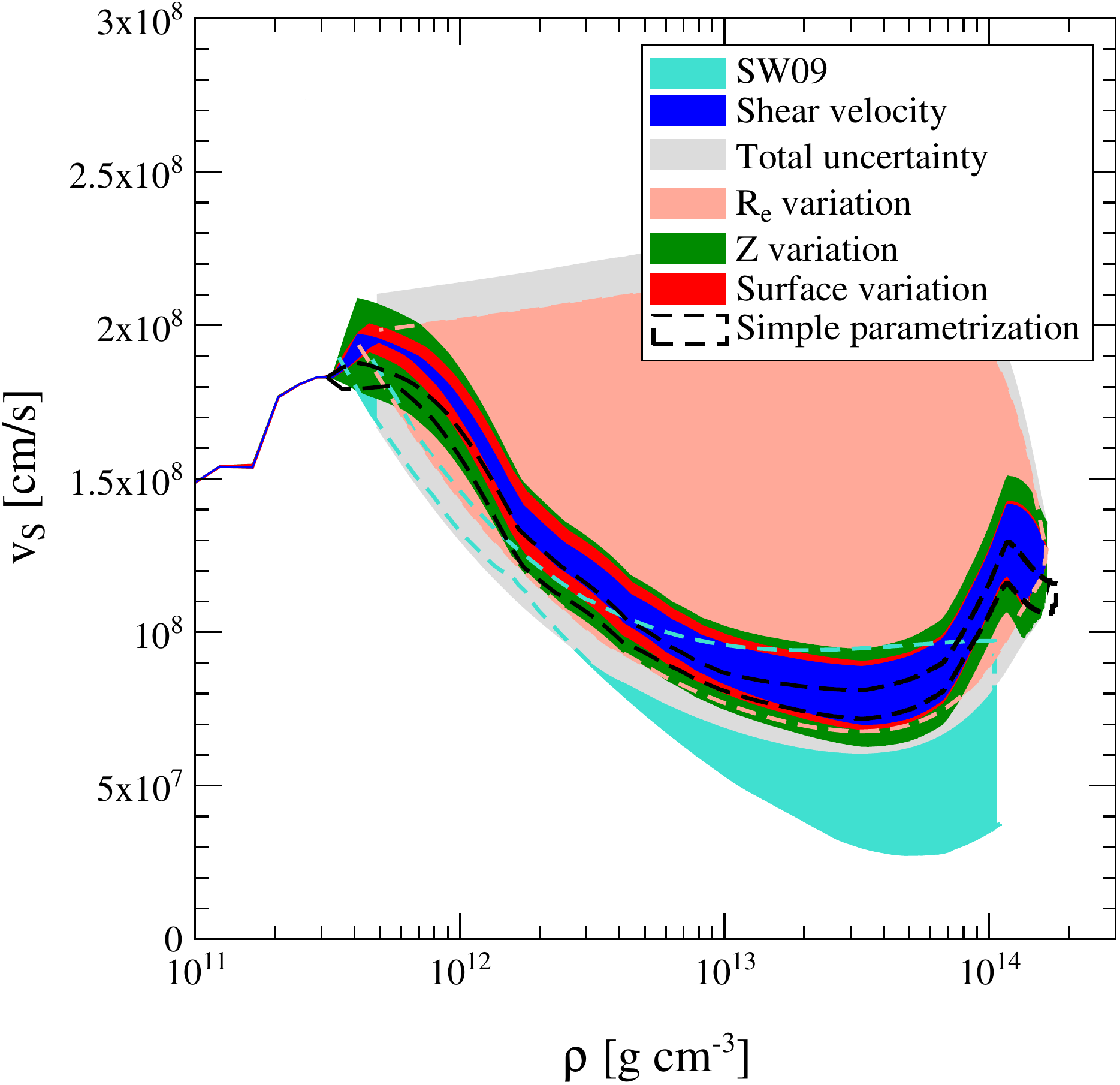}
\caption{Shear velocities as a function of density. In addition to the 
uncertainty bands from Fig.~\ref{fig:munmup}, I show the uncertainty in 
the entrainment parameter $R_e$ (light-red band) as well as the combined 
total uncertainty (grey band). I compare my results with the velocity band 
from Ref.~\cite{SteinerWatts2009} (SW09), which used various Skyrme 
interactions as input.
 \label{fig:ShearSpeed}}
\end{figure} 

\begin{figure}[t]
\centering
\includegraphics[trim= 0 0 0 0cm, clip=,width=0.44\textwidth]{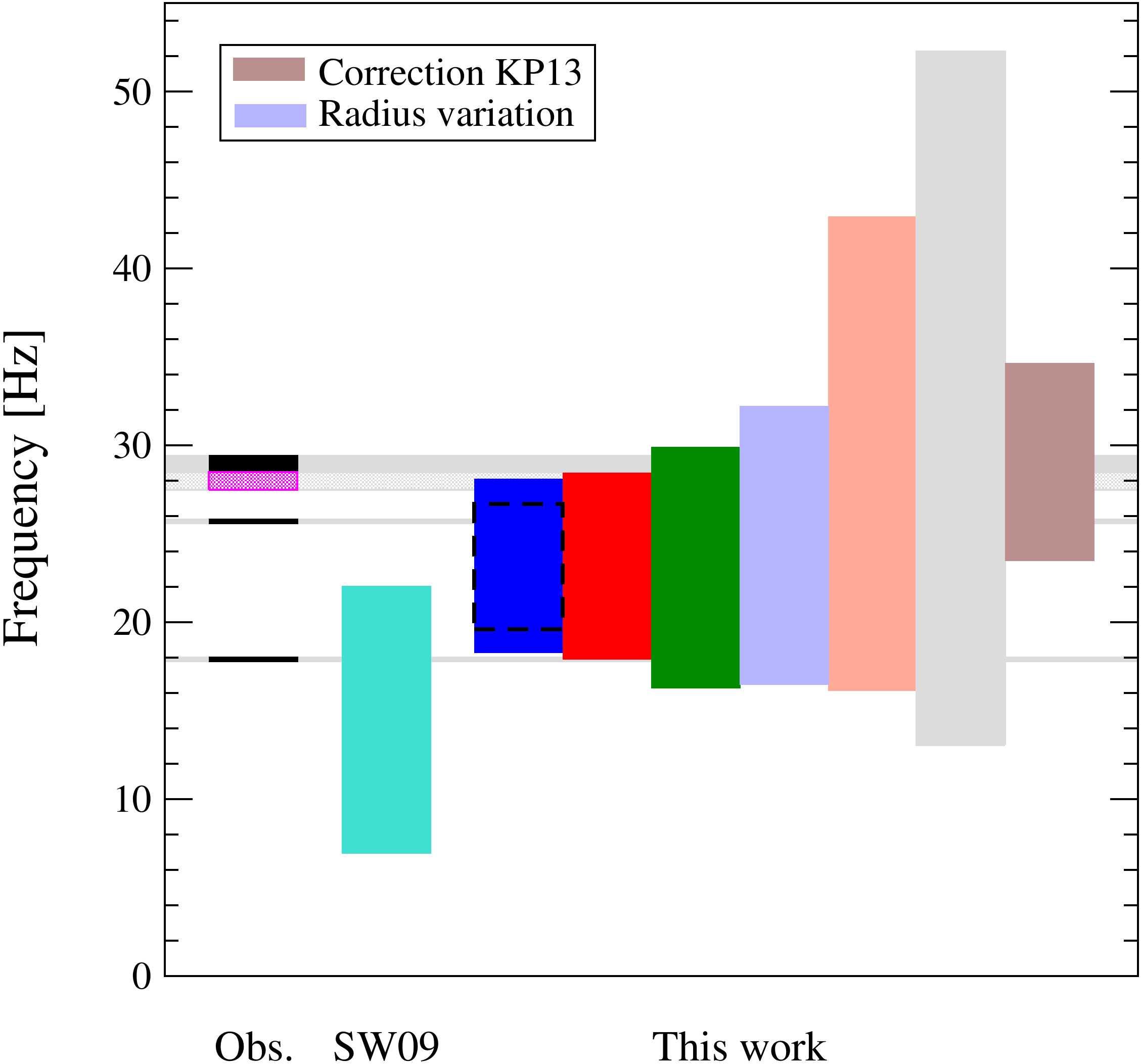}
\caption{Frequencies of the fundamental ($n=0, l=2$) shear mode for 
the shear velocities of Fig.~\ref{fig:ShearSpeed}. All bands include a 
variation of the total neutron-star mass from $M=1.00 -1.97 \,M_{\odot}$ 
as described in the text. In addition, I show the frequency range when 
varying the neutron-star radius according to the uncertainties from 
Ref.~\cite{Hebeler:2010b} (light-blue band) and the frequencies when
correcting $R_e$ by the suggested factor by Kobyakov and 
Pethick~\cite{Kobyakov:2013} (KP13, brown band). I compare with the results 
of Ref.~\cite{SteinerWatts2009} (SW09) and with observed QPO frequencies
from SGR~1806-20 (black lines) and SGR~1900+14 (purple lines).
\label{fig:FundFreq}}
\end{figure} 

Three such giant bursts have been detected so far. In their 
afterglow, in addition to the modulation associated with the magnetar's 
rotation, several quasiperiodic oscillations (QPOs) in the frequency range 
from 18-1800 Hz were observed~\cite{Israel:2005, Strohmayer:2005,
Watts:2006,Strohmayer:2006, Hambaryan:2010}. These oscillations 
have been interpreted as torsional shear modes of the 
crust~\cite{Duncan:1998my}. Crustal shear modes have lower excitation 
energies compared to other vibrational modes, and their excitation is 
plausible considering typical energy releases in giant flares~\cite{Lander:2014,
Piro:2005}. Furthermore, models of torsional shear modes are in qualitative
agreement with many of the observed QPO frequencies and their scaling 
behavior. 

If this interpretation were correct, QPOs could be used to infer crustal 
shear properties as well as to put constraints on the mass-radius relation of 
neutron stars. The latter is possible because different oscillation modes have 
different dependencies on neutron star radius, crust thickness, and 
mass~\cite{Samuelsson:2006tt}. A description of QPOs solely from crustal
properties makes use of the free-slip boundary condition, meaning that crust
and core can be treated independently. Because the strong magnetic fields 
in the magnetars couple crust and core, this approximation seems to be 
inaccurate in general~\cite{Levin:2006ck, Gabler:2010rp}, and QPOs are 
most likely global oscillation modes. 

Initial studies of global magnetar oscillations~\cite{Glampedakis:2006, 
Levin:2006qd} found that all crustal modes will couple with and transfer 
energy to the neutron-star core on a very short time scale, creating Alfv\'en
continua. Further studies~\cite{Colaiuda:2010pc, vanHoven:2010gy} 
found that the appearing spectrum is very rich. There may appear continua of 
global oscillation modes with strong signals at their endpoints. Moreover, the 
star may permit discrete core Alfv\'en modes, which are possible probes of 
the neutron-star core. Last, discrete strong crust-dominated modes may 
appear, which can be very close to the pure crustal modes. If these 
crust-dominated modes lie within the continua, they will be absorbed and 
disappear from the spectrum. If these modes, instead, lie in the gaps 
between the continua, they will be very strong. In more recent studies of
global magneto-elastic oscillations for different 
magnetic field geometries and strengths, however, it was found that no 
crust-dominated shear modes will survive in the QPO 
spectra~\cite{Gabler:2010rp, Gabler:2011am, Gabler:2012jh,
Gabler:2016rth, Passamonti:2016jfo}. The authors excluded crustal
shear oscillations as an explanation of observed magnetar QPOs. 
Magnetic-field effects, thus, complicate the correct identification of 
observed QPO frequencies in terms of different oscillation modes, 
which is a challenging and open problem.

Although crustal shear oscillations may not be sufficient to explain the 
observed QPO frequencies, in this work I will assume that at least some 
QPO frequencies can be described by pure crustal oscillation modes 
(which may be close to global crust-dominated shear modes). 
Then the fundamental frequency is found to be comparable in size to 
the lowest observed QPO frequencies: 18, 26, and 29 Hz for the hyperflare 
SGR~1806-20~\cite{Strohmayer:2006}, and 28 Hz for 
SGR~1900+14~\cite{Strohmayer:2005}. For example, in 
Ref.~\cite{SteinerWatts2009}, the calculated fundamental shear mode 
frequencies ranged from $7$ to $22$~Hz and were compatible with the 18 Hz 
QPO. In Ref.~\cite{Deibel:2013sia}, the fundamental shear mode was 
matched either with the 18~Hz QPO or the 28~Hz QPO, based on the 
employed EOS model.

In this section, I calculate the frequency of the fundamental crustal 
shear mode ($n=0, l=2$) with free-slip boundary conditions based on 
the results for the inner-crust shear properties. Because I treat crust 
and core separately, I can parametrize the core by its mass and radius
and solve the Tolman-Oppenheimer-Volkoff equations starting at the
crust-core boundary. 
I will vary the core mass $M_C$ so that the total neutron-star mass 
range is $M=1.00 -1.97 \, M_{\odot}$, but I expect the mass 
variation to have only a small effect on the frequencies of the $n=0$ 
modes~\cite{SteinerWatts2009}. For each core mass, I choose 
the corresponding core radius $R_C$ in such a way that the total 
neutron-star radius coincides with the mean value of the mass-radius 
band of Ref.~\cite{Hebeler:2013}. This band was obtained using the 
same chiral interactions as used in this work and, thus, is consistent 
with my approach. It was constrained only by causality and the
observation of $2 M_{\odot}$ neutron stars. To include the uncertainty 
in the mass-radius band, I will vary the neutron-star radii within its 
boundaries. Doing this, I obtain a neutron-star crust thickness of $0.64
-1.37 \km$ for a typical $1.4 M_{\odot}$ neutron star for the crust EOS
of this work, in very good agreement with the findings of 
Ref.~\cite{Steiner:2014pda}.

I use the Newtonian perturbation model employed in Ref.~\cite{Piro:2005} 
with the $l$ scaling of Ref.~\cite{SteinerWatts2009}, leading to the 
perturbation equation for toroidal shear modes,
\begin{align}
\frac{(\mu \xi')'}{\rho_C}+v_A^2\xi''+\left(\omega^2\left(1+\frac{v_A^2}{c^2} \right)-\frac{(l^2+l-2)\mu}{\rho_C R^2}\right) \xi =0\,, \label{eq:perteqn}
\end{align}
where $\xi$ is the displacement, which has only a horizontal component,
primes correspond to derivatives with respect to the vertical direction, 
$v_A=B/(4 \pi \rho_C)^{1/2}$ is the Alfv\'en speed with the magnetic field
$B$, $\omega=2\pi f$ the frequency of the shear mode, and $R$ the 
neutron-star radius. I vary $B=10^{14}-10^{15}$G, but for the 
fundamental mode the influence of the magnetic field is negligible and one 
recovers the same frequencies when using $B=0$. The 
first radial overtone, however, changes for field strengths above $10^{15} 
$G~\cite{Piro:2005, Sotani:2006at, Colaiuda:2010pc, Deibel:2013sia}.

I solve the perturbation equation~\eqref{eq:perteqn} with the following
boundary conditions. At the ocean/crust interface, at densities of 
$\approx 5\times 10^7 \text{g cm}^{-3}$, I assume no horizontal shear 
stress or traction, $\xi'(R)=0$. I then vary the frequency $\omega$ until 
I find a solution with no horizontal shear stress or traction at the 
crust-core boundary, $\xi'(R_C)=0$ (because crust and core are decoupled).
I finally correct the obtained frequency for gravitational redshift,
\begin{align}
f_{\text{obs}}=f_{\text{emit}}\sqrt{1-\frac{r_S}{R}}\,,
\end{align}
with the Schwarzschild radius $r_S=2 G M/c^2$. The obtained frequencies 
will depend on the crust thickness $\Delta R$, the neutron-star mass $M$ 
and radius $R$.

I present the results for the fundamental crustal shear mode in 
Fig.~\ref{fig:FundFreq}. I show bands for all considered sources of 
uncertainty (each band also includes the EOS uncertainties and the 
mass variation), and compare with the lowest observed QPO frequencies 
as well as with the results of Ref.~\cite{SteinerWatts2009}. The latter 
range from $7$ to $22$~Hz and include a neutron-star mass variation 
from $1.2M_{\odot}$ to $1.97 M_{\odot}$. These results are only compatible with 
the 18~Hz QPO from SGR~1806-20. When setting $R_e=1$, as in 
Ref.~\cite{SteinerWatts2009}, I obtain frequencies of $16.2-23.5$
Hz for a mass variation in the same range. Both calculations, thus, are 
in very good agreement, and consistent with the shear speeds of 
Fig.~\ref{fig:ShearSpeed}. Furthermore, these results also agree very 
well with the calculation of Ref.~\cite{Sotani:2006at} for the fundamental 
oscillation mode.

Considering superfluid neutrons and entrainment effects, the dynamical 
mass density decreases and the shear velocities increase. This also leads
to an increase of the fundamental frequencies. For a higher entrainment 
coefficient $R_e$ the frequencies are lower, while less entrainment leads 
to higher frequencies. Using the $R_e$ values from 
Ref.~\cite{Chamel:2012zn}, I obtain a frequency band of $18.3-28.1$
Hz for a mass variation of $M=1.00 -1.97 \, M_{\odot}$ (blue band),
approximately $10\%$ higher than for neglecting entrainment. This is 
consistent with the findings of Ref.~\cite{Samuelsson:2009xz}. Roughly 
$50\%$ of the uncertainty stems from the mass variation, and $\approx 
50\%$ from the uncertainty in the crust EOS. 

\begin{figure}[t]
\centering
\includegraphics[trim= 0 0 0 0cm, clip=,width=0.44\textwidth]{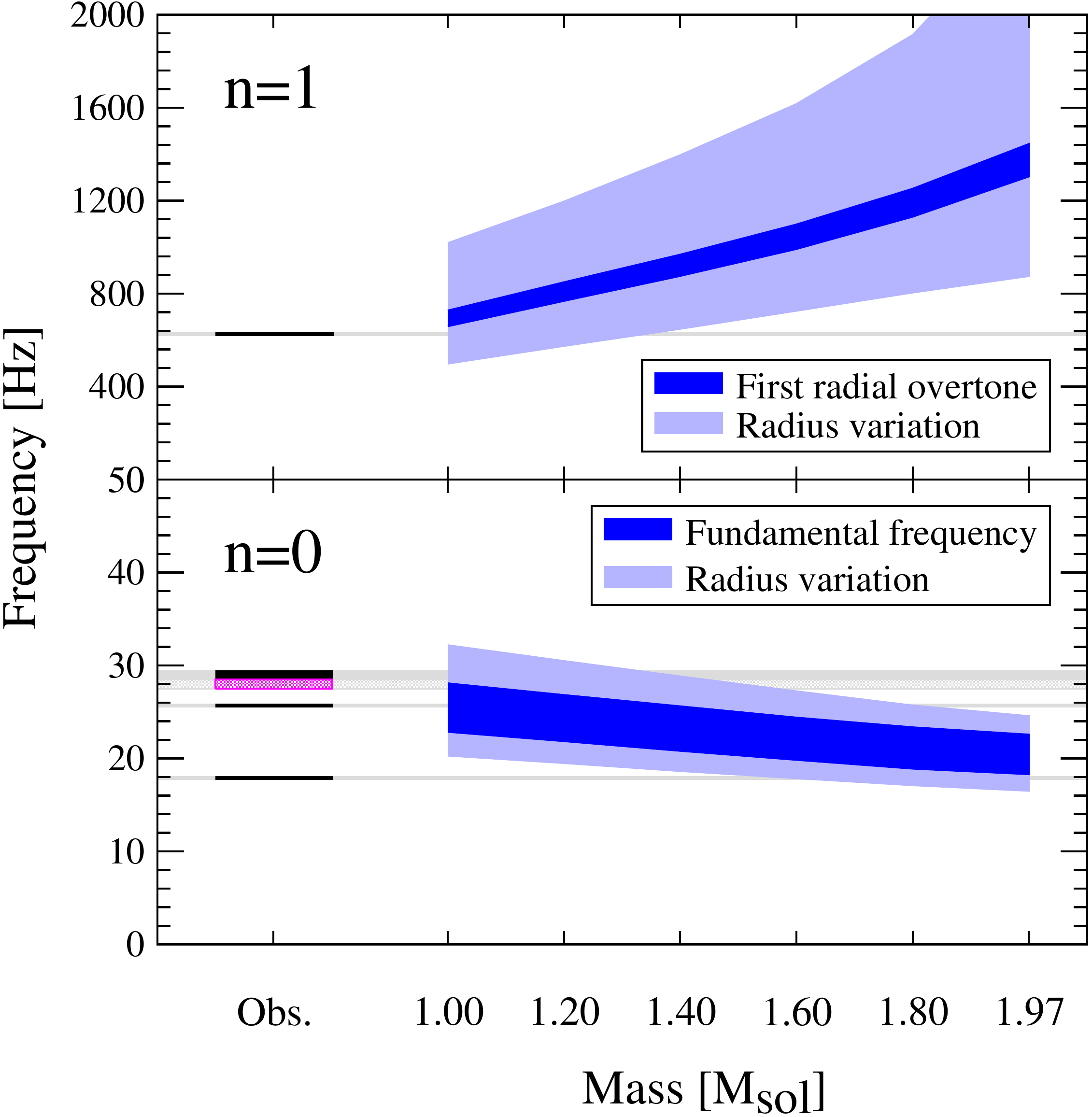}
\caption{Lower panel: Frequencies of the $n=0, l=2$ fundamental shear 
mode for $R_e$ from Ref.~\cite{Chamel:2012zn} and various neutron-star 
masses (blue band). I also show the uncertainties from radius variation 
(light-blue band), and compare to the lowest observed QPO frequencies 
from SGR~1806-20 (black lines) and SGR~1900+14 (purple line). Upper 
panel: The same for the $n=1$ first radial overtone compared to 
the observed $626$ Hz QPO. 
\label{fig:Freqn0}}
\end{figure}

I show the dependence of the fundamental frequency on the different 
neutron star masses in the lower panel of Fig.~\ref{fig:Freqn0}. For the 
lightest neutron star I obtain a frequency range of $22.9-28.1$~Hz, and
for the heaviest neutron star I obtain a range of $18.3-22.6$~Hz: Lower 
mass stars will have slightly higher shear frequencies. These frequencies 
are compatible with the observed $28-29$~Hz QPOs for a neutron-star mass
around $1.0 M_{\odot}$. For a typical $1.4 \, M_{\odot}$ neutron star, 
I find a frequency range of $20.8-25.6$~Hz, in very good agreement with
the findings of Ref.~\cite{Sotani:2013jya} for the $L$ range of chiral 
interactions . I also show the uncertainty due 
to radius variation for same-mass neutron stars (light-blue band). This 
band reflects the uncertainty of the core EOS. For each individual neutron 
star, the uncertainties due the crust EOS and the core EOS are of similar 
size. Including the radius variation, the uncertainty band for the fundamental 
shear mode grows to $16.5-32.2$~Hz; see also Fig.~\ref{fig:FundFreq}. This
range is compatible with the $28$ Hz QPO for stars with masses $M\leq 1.4
M_{\odot}$ and with the $18$ Hz QPO for heavy neutron stars with 
$M\geq 1.6 M_{\odot}$. 

Contrary to the EOS uncertainties, the uncertainties in the crust modeling
affect the frequency range only modestly. Compared to the original band
(without radius variation, $18.3-28.1$~Hz), the effect due to different surface
parameters is almost negligible, and the range increases to $18.0-28.4$~Hz. 
The variation of the proton number of the crust nuclei, $Z$, increases the
uncertainty range mildly to $16.3-29.9$~Hz. This behavior is consistent with 
the shear velocities of Fig.~\ref{fig:ShearSpeed}. The variation of $R_e$ 
has a sizable impact on the results, which is already clear from the shear
velocities. A variation of $R_e$ from full to no entrainment leads to an 
uncertainty band of $16.2-42.9$ Hz. Entrainment, thus, is a major source 
of uncertainty, in addition to the EOS uncertainty.

\begin{figure}[t]
\centering
\includegraphics[trim= 0 0 0 0cm, clip=,width=0.45\textwidth]{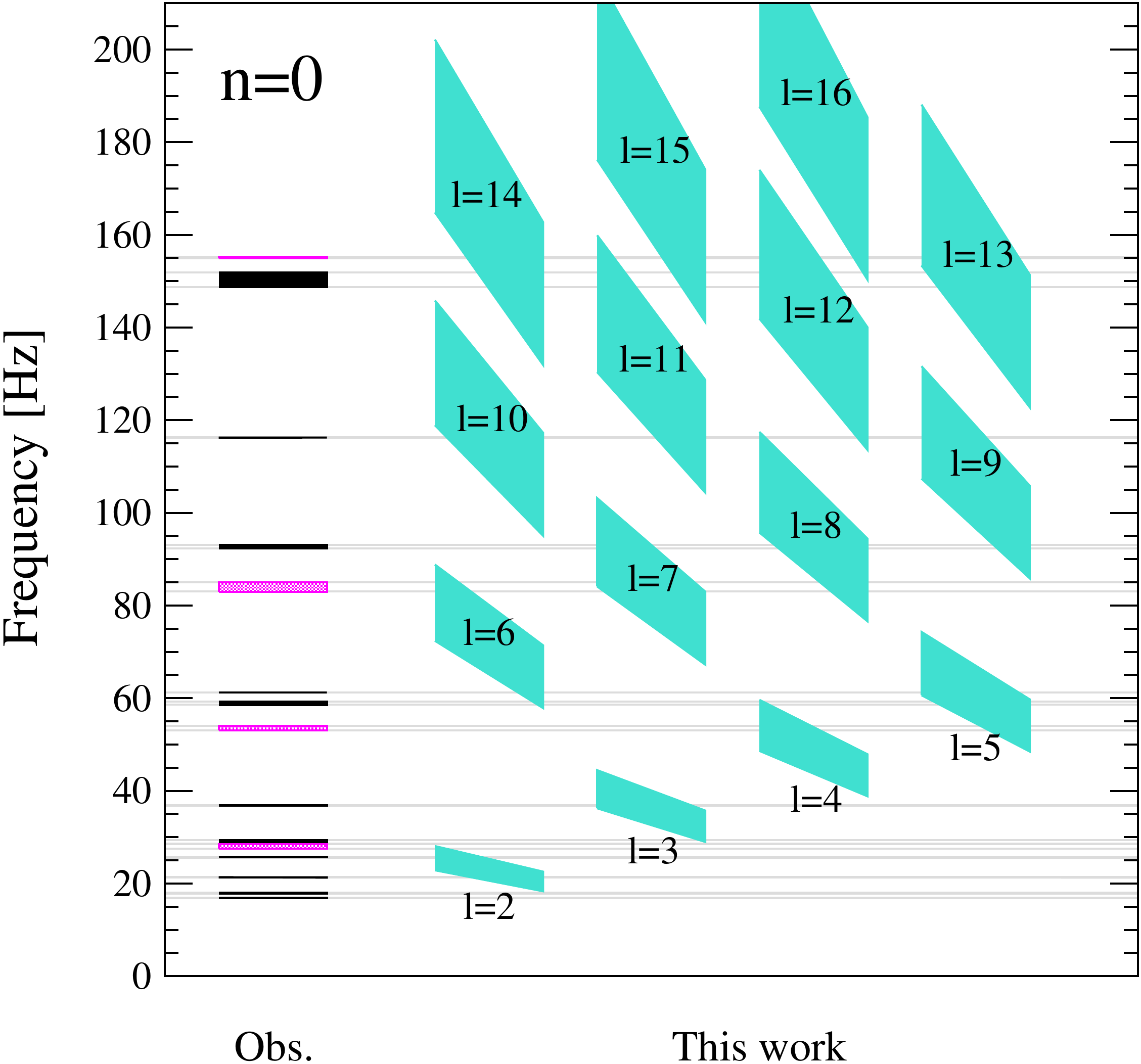}
\caption{Frequencies of the $n=0$ modes for $2\leq l \leq 16$ for $R_e$ 
from Ref.~\cite{Chamel:2012zn}. For each $l$, I show the frequencies
obtained for a $1.00 M_{\odot}$ neutron star (left boundary) up to a 
$1.97 M_{\odot}$ neutron star (right boundary). I compare with observed
QPO frequencies from SGR~1806-20 (black lines) and SGR~1900+14 (purple
lines).
\label{fig:FreqL}}
\end{figure} 

The blue frequency band (using $R_e$ values of Ref.~\cite{Chamel:2012zn})
is only compatible with any observed QPO frequency for very light or very 
heavy neutron stars. Kobyakov and Pethick~\cite{Kobyakov:2013} suggested 
that lattice
vibrations can reduce $R_e$, and, in the relevant density regime, they
found a correction factor of $\approx 0.4$. Multiplying $R_e$ by
a constant correction factor of this magnitude (which is a good approximation 
in the density range $n=0.01-0.06 \fm^{-3}$) leads to a frequency
band of $23.5-34.6$~Hz, which is in very good agreement with the observed 
$28-29$~Hz QPO frequencies for all neutron-star masses. Additional effects 
may influence the fraction of entrained neutrons, e.g., neutron pairing, and 
a better understanding of neutron entrainment is a necessary condition for
obtaining neutron-star properties from asteroseismology. 

Combining all sources of uncertainty, I find the lowest fundamental pure 
crustal shear frequency, which is consistent with current nuclear-physics 
constraints, to be $13.1$~Hz, while the highest frequency is $52.3$ Hz. 
Current uncertainties lead to a sizable frequency range. This range may 
additionally increase if nuclear pasta phases are 
considered~\cite{Sotani:2011nn, Passamonti:2016jfo}. For the frequency of 
the fundamental mode, e.g., Passamonti and Pons.~\cite{Passamonti:2016jfo} 
found a reduction 
of up to $40 \%$ when pasta phases are included, while the effect is weaker 
for higher-frequency shear modes. Nevertheless, if one were to match an 
observed QPO frequency with the fundamental crustal shear mode, then 
based on my calculations an identification with the $28-29$~Hz QPO seems
to be likely. 

Finally, for the simple parametrization and the same parameters chosen to 
obtain the blue band, I find a frequency range of $19.6-26.7$~Hz; see 
Fig.~\ref{fig:FundFreq}. The two parametrizations are in excellent agreement
because the fundamental mode frequency is mainly sensitive to the shear 
velocities at higher densities, where both parametrizations agree.
 
\section{Higher shear modes} \label{sec:n1}

In addition to the lowest observed QPO frequencies of $18-29$ Hz, several 
other frequencies up to $1800$ Hz have been observed. A large number of 
them, with frequencies up to $\approx 160~\text{Hz}$, may be identified 
with higher-$l$ overtones of the fundamental $n=0$ mode, while the QPO 
at $624$ Hz was identified as the first radial $n=1$ overtone. However, 
for the latter, this identification is not clear because crustal shear
modes at such high frequencies are strongly damped; see, e.g., 
Ref.~\cite{vanHoven:2010gy}.

In Fig.~\ref{fig:FreqL}, I show the observed frequencies up to 160 Hz 
from Refs.~\cite{Israel:2005,Strohmayer:2005,Strohmayer:2006,
Hambaryan:2010} and compare these to calculated pure crustal oscillation
frequencies for $n=0$ and $2\leq l \leq 16$. For each oscillation mode, 
I show the frequency variation for different masses ranging from 
$1.0 M_{\odot}$ to $1.97 M_{\odot}$, similar to Fig.~\ref{fig:Freqn0}. For every given 
mass, every observed frequency above 60 Hz could be matched with a 
certain crustal shear mode within nuclear physics uncertainties. One finds 
gaps only between the lowest shear modes with $l=2, 3, 4, 5$. At higher
frequencies, due to the larger uncertainties, bands for different modes 
start to overlap and could be identified with the same observed QPO 
frequency. A possible mode assignment, thus, is ambiguous. The results
are in good agreement with the results of Ref.~\cite{Sotani:2013jya} 
for the chiral range of $L$ values and with the results for torsional 
crustal shear modes including entrainment of Ref.~\cite{Passamonti:2016jfo}.

Nevertheless, if one were to match observed QPO frequencies with pure
crustal modes and identify the $28-29$ Hz QPO mode as the fundamental 
$n=0, l=2$ frequency, then both sources SGR~1806-20 and SGR~1900+14 
seem to be lower-mass neutron stars with $M < 1.4 M_{\odot}$, with 
SGR~1900+14 being heavier than SGR~1806-20. The observed QPOs 
below 100~Hz then could be 
identified with $l=4$ and $l=6$ modes, while the QPO at 37~Hz does 
not seem to correspond to any crustal oscillation mode. The 
parameter $R_e$, though, will have a strong impact on all calculated 
frequencies, and a possible reduction of $R_e$, as suggested in 
Ref.~\cite{Kobyakov:2013}, will move all bands up, as shown in the 
previous section.

I now turn to the $n=1$ radial overtone. In Fig.~\ref{fig:Freqn0}, I 
show results for the $n=1$ overtone for different neutron stars with 
masses ranging from $1.0 M_{\odot}$ to $1.97 M_{\odot}$. For a $1.0 M_{\odot}$ 
neutron star, I obtain a frequency range of $660-727$~Hz, while for the 
$1.97 M_{\odot}$ neutron star I obtain a range of $1305-1443$~Hz. 
The mass dependence of the first radial overtone can easily be understood
because this mode strongly depends on the crust thickness. Heavier 
neutron stars have thinner crusts than lower-mass neutron stars and 
thinner crusts require higher frequencies to fulfill the boundary conditions.  
If the observed 624~Hz QPO is identified with the first radial overtone, 
this suggests that SGR~1806-20 is a lower-mass neutron star, which is 
consistent with the previous results for the $n=0$ shear modes. 

I also show the uncertainties when varying the neutron-star radius, as 
before. This has a sizable effect on the first radial overtone and lowering 
the radius for a $1.0 M_{\odot}$ neutron star can increase the frequency 
up to 1017~Hz. Increasing the radius can decrease the frequency to 500~Hz. 
This can again be understood in terms of crust thickness: For a neutron star 
of a given mass, smaller (larger) radii lead to thinner (thicker) crusts. For the
heaviest neutron star, I find a total frequency range of $876-2436$~Hz when
varying the radius. 

These results lead to a total uncertainty band for the first radial overtone 
of $500-2436$~Hz originating only in the EOS and mass uncertainty. When
additionally varying $R_e$, the uncertainty increases to $454-4136$~Hz. 
This illustrates that QPOs can in principle be a powerful tool to infer properties 
of neutron stars and/or constrain the EOS of nuclear matter but within 
current uncertainties a mode identification is not possible. Considering global
oscillation modes is likely to increase the uncertainty which makes mode
identification even more difficult. Additional information is needed to make 
robust predictions. An improved determination of $R_e$ with small 
uncertainties would be a very useful first step. 
 
\section{Summary and Outlook}\label{sec:outlook}

In this paper I have studied the influence of different sources of 
uncertainty on the spectrum of shear modes in the neutron-star inner crust. 
To capture the uncertainties in the nuclear interactions I used 
parametrizations for the energy per particle of nuclear matter, which were 
fit to chiral EFT calculations of pure neutron matter and the empirical 
saturation point. Using these interactions, I modeled the inner crust in 
the Wigner-Seitz approximation and studied the impact of uncertainties in 
the crust composition and the surface energy parameters. While the 
uncertainty in the crust composition has the largest impact on the geometry 
of the Wigner-Seitz cell, the uncertainty in the inner-crust EOS is dominated 
by the nuclear interactions.

Using the inner-crust model, I determined the shear modulus and shear 
velocities in the neutron star crust with uncertainties. For the shear modulus
I found that the main uncertainty stems from the crust composition at 
low densities and from the neutron-matter EOS at higher 
densities. For the shear velocities, the main source of uncertainty is neutron
entrainment, which leads to a variation up to factor of 2 in the 
neutron-star crust.

Using free-slip boundary conditions, I calculated the frequencies of the
fundamental crustal shear modes and compared the calculation to observed 
QPO frequencies. I obtained fundamental frequencies ranging from $18$ to 
$28$ Hz, with a total uncertainty band of $13-52$~Hz. I identified three major
sources of uncertainty: first, the EOS of nuclear matter up to saturation 
density, which sets the inner-crust EOS; second, the EOS above saturation 
density, which enters the calculation via the radius variation; and third, the
entrainment factor. The effect of the uncertainties in the crust composition 
and surface parameters on the shear-mode frequencies, instead, are small. 
Both an improved description of the EOS with reduced theoretical
uncertainties and a better determination of the entrainment factor are 
necessary to reliably model crustal shear oscillations. Corrections to the 
value of entrainment, as suggested in Ref.~\cite{Kobyakov:2013}, lower 
the number of neutrons locked in the lattice and lead to an increase of the
calulated fundamental frequencies to $24-35$ Hz. If the fundamental QPO
frequencies can be described in terms of crustal shear modes, an 
identification of the fundamental shear mode with the $28-29$ Hz QPO, thus, 
seems to be likely.

I also performed calculations of oscillation modes with $n=0$ and 
$2 \leq l \leq 16$ as well as of the $n=1, l=2$ mode. These calculations 
are very dependent on the neutron-star parameters, but show that every 
observed QPO frequency could be described by at least one crustal shear 
mode within uncertainties. While QPOs in principle could be used to infer 
neutron-star properties, current uncertainties are quite sizable and hinder 
the clear identification of modes, which, in turn, impedes the extraction 
of robust constraints. The computation of shear modes in the neutron-star 
crust would mostly benefit from a reduction of (a) the uncertainty of the 
EOS of neutron matter at densities below and above saturation density, 
and (b) a determination of the entrainment factor with robust theoretical
uncertainties. In addition, the influence of nuclear pasta phases has to be
investigated in detail. Together with a global neutron-star oscillation model, 
which properly includes the effects of the strong magnetic fields, these
improvements would allow comparisons with observed frequencies to 
reliably identify modes and infer properties of neutron stars. On the other 
hand, additional information on the QPO sources, e.g., masses, would allow 
one to put constraints on the EOS or the entrainment factor.  

\begin{acknowledgments}
The author thanks Sanjay Reddy and Achim Schwenk for valuable input and
feedback on the manuscript. The author also thanks Nicolas Chamel, Christian
Drischler, Dmitry Kobyakov, and Anna Watts for useful discussions. This work 
was supported by the National Science Foundation under Grant No. 
PHY-1430152 (JINA Center for the Evolution of the Elements) and by the 
US DOE Grant No. DE-FG02-00ER41132.
\end{acknowledgments}

\bibliographystyle{apsrev4-1} 
\bibliography{bibliography}{}

\begin{thebibliography}{81}%
\makeatletter
\providecommand \@ifxundefined [1]{%
 \@ifx{#1\undefined}
}%
\providecommand \@ifnum [1]{%
 \ifnum #1\expandafter \@firstoftwo
 \else \expandafter \@secondoftwo
 \fi
}%
\providecommand \@ifx [1]{%
 \ifx #1\expandafter \@firstoftwo
 \else \expandafter \@secondoftwo
 \fi
}%
\providecommand \natexlab [1]{#1}%
\providecommand \enquote  [1]{``#1''}%
\providecommand \bibnamefont  [1]{#1}%
\providecommand \bibfnamefont [1]{#1}%
\providecommand \citenamefont [1]{#1}%
\providecommand \href@noop [0]{\@secondoftwo}%
\providecommand \href [0]{\begingroup \@sanitize@url \@href}%
\providecommand \@href[1]{\@@startlink{#1}\@@href}%
\providecommand \@@href[1]{\endgroup#1\@@endlink}%
\providecommand \@sanitize@url [0]{\catcode `\\12\catcode `\$12\catcode
  `\&12\catcode `\#12\catcode `\^12\catcode `\_12\catcode `\%12\relax}%
\providecommand \@@startlink[1]{}%
\providecommand \@@endlink[0]{}%
\providecommand \url  [0]{\begingroup\@sanitize@url \@url }%
\providecommand \@url [1]{\endgroup\@href {#1}{\urlprefix }}%
\providecommand \urlprefix  [0]{URL }%
\providecommand \Eprint [0]{\href }%
\providecommand \doibase [0]{http://dx.doi.org/}%
\providecommand \selectlanguage [0]{\@gobble}%
\providecommand \bibinfo  [0]{\@secondoftwo}%
\providecommand \bibfield  [0]{\@secondoftwo}%
\providecommand \translation [1]{[#1]}%
\providecommand \BibitemOpen [0]{}%
\providecommand \bibitemStop [0]{}%
\providecommand \bibitemNoStop [0]{.\EOS\space}%
\providecommand \EOS [0]{\spacefactor3000\relax}%
\providecommand \BibitemShut  [1]{\csname bibitem#1\endcsname}%
\let\auto@bib@innerbib\@empty
\bibitem [{\citenamefont {Demorest}\ \emph {et~al.}(2010)\citenamefont
  {Demorest}, \citenamefont {Pennucci}, \citenamefont {Ransom}, \citenamefont
  {Roberts},\ and\ \citenamefont {Hessels}}]{Demorest2010}%
  \BibitemOpen
  \bibfield  {author} {\bibinfo {author} {\bibfnamefont {P.}~\bibnamefont
  {Demorest}}, \bibinfo {author} {\bibfnamefont {T.}~\bibnamefont {Pennucci}},
  \bibinfo {author} {\bibfnamefont {S.}~\bibnamefont {Ransom}}, \bibinfo
  {author} {\bibfnamefont {M.}~\bibnamefont {Roberts}}, \ and\ \bibinfo
  {author} {\bibfnamefont {J.}~\bibnamefont {Hessels}},\ }\href {\doibase
  10.1038/nature09466} {\bibfield  {journal} {\bibinfo  {journal} {Nature}\
  }\textbf {\bibinfo {volume} {467}},\ \bibinfo {pages} {1081} (\bibinfo {year}
  {2010})},\ \Eprint {http://arxiv.org/abs/1010.5788} {arXiv:1010.5788
  [astro-ph.HE]} \BibitemShut {NoStop}%
\bibitem [{\citenamefont {Antoniadis}\ \emph {et~al.}(2013)\citenamefont
  {Antoniadis}, \citenamefont {Freire}, \citenamefont {Wex}, \citenamefont
  {Tauris}, \citenamefont {Lynch} \emph {et~al.}}]{Antoniadis2013}%
  \BibitemOpen
  \bibfield  {author} {\bibinfo {author} {\bibfnamefont {J.}~\bibnamefont
  {Antoniadis}}, \bibinfo {author} {\bibfnamefont {P.~C.}\ \bibnamefont
  {Freire}}, \bibinfo {author} {\bibfnamefont {N.}~\bibnamefont {Wex}},
  \bibinfo {author} {\bibfnamefont {T.~M.}\ \bibnamefont {Tauris}}, \bibinfo
  {author} {\bibfnamefont {R.~S.}\ \bibnamefont {Lynch}},  \emph {et~al.},\
  }\href {\doibase 10.1126/science.1233232} {\bibfield  {journal} {\bibinfo
  {journal} {Science}\ }\textbf {\bibinfo {volume} {340}},\ \bibinfo {pages}
  {6131} (\bibinfo {year} {2013})},\ \Eprint {http://arxiv.org/abs/1304.6875}
  {arXiv:1304.6875 [astro-ph.HE]} \BibitemShut {NoStop}%
\bibitem [{\citenamefont {Lattimer}(2014)}]{Lattimer:2014}%
  \BibitemOpen
  \bibfield  {author} {\bibinfo {author} {\bibfnamefont {J.~M.}\ \bibnamefont
  {Lattimer}},\ }\href {\doibase 10.1007/s10714-014-1713-3} {\bibfield
  {journal} {\bibinfo  {journal} {Gen. Rel. Grav.}\ }\textbf {\bibinfo {volume}
  {46}},\ \bibinfo {pages} {1713} (\bibinfo {year} {2014})}\BibitemShut
  {NoStop}%
\bibitem [{\citenamefont {Hebeler}\ \emph {et~al.}(2013)\citenamefont
  {Hebeler}, \citenamefont {Lattimer}, \citenamefont {Pethick},\ and\
  \citenamefont {Schwenk}}]{Hebeler:2013}%
  \BibitemOpen
  \bibfield  {author} {\bibinfo {author} {\bibfnamefont {K.}~\bibnamefont
  {Hebeler}}, \bibinfo {author} {\bibfnamefont {J.~M.}\ \bibnamefont
  {Lattimer}}, \bibinfo {author} {\bibfnamefont {C.~J.}\ \bibnamefont
  {Pethick}}, \ and\ \bibinfo {author} {\bibfnamefont {A.}~\bibnamefont
  {Schwenk}},\ }\href {\doibase 10.1088/0004-637X/773/1/11} {\bibfield
  {journal} {\bibinfo  {journal} {Astrophys. J.}\ }\textbf {\bibinfo {volume}
  {773}},\ \bibinfo {pages} {11} (\bibinfo {year} {2013})},\ \Eprint
  {http://arxiv.org/abs/1303.4662} {arXiv:1303.4662 [astro-ph.SR]} \BibitemShut
  {NoStop}%
\bibitem [{\citenamefont {Watts}\ \emph {et~al.}(2016)\citenamefont {Watts}
  \emph {et~al.}}]{Watts:2016uzu}%
  \BibitemOpen
  \bibfield  {author} {\bibinfo {author} {\bibfnamefont {A.~L.}\ \bibnamefont
  {Watts}} \emph {et~al.},\ }\href {\doibase 10.1103/RevModPhys.88.021001}
  {\bibfield  {journal} {\bibinfo  {journal} {Rev. Mod. Phys.}\ }\textbf
  {\bibinfo {volume} {88}},\ \bibinfo {pages} {021001} (\bibinfo {year}
  {2016})},\ \Eprint {http://arxiv.org/abs/1602.01081} {arXiv:1602.01081
  [astro-ph.HE]} \BibitemShut {NoStop}%
\bibitem [{\citenamefont {Schneider}\ \emph {et~al.}(2013)\citenamefont
  {Schneider}, \citenamefont {Horowitz}, \citenamefont {Hughto},\ and\
  \citenamefont {Berry}}]{Schneider:2013dwa}%
  \BibitemOpen
  \bibfield  {author} {\bibinfo {author} {\bibfnamefont {A.~S.}\ \bibnamefont
  {Schneider}}, \bibinfo {author} {\bibfnamefont {C.~J.}\ \bibnamefont
  {Horowitz}}, \bibinfo {author} {\bibfnamefont {J.}~\bibnamefont {Hughto}}, \
  and\ \bibinfo {author} {\bibfnamefont {D.~K.}\ \bibnamefont {Berry}},\ }\href
  {\doibase 10.1103/PhysRevC.88.065807} {\bibfield  {journal} {\bibinfo
  {journal} {Phys. Rev.}\ }\textbf {\bibinfo {volume} {C88}},\ \bibinfo {pages}
  {065807} (\bibinfo {year} {2013})},\ \Eprint {http://arxiv.org/abs/1307.1678}
  {arXiv:1307.1678 [nucl-th]} \BibitemShut {NoStop}%
\bibitem [{\citenamefont {Chamel}\ and\ \citenamefont
  {Haensel}(2008)}]{Chamel:2008ca}%
  \BibitemOpen
  \bibfield  {author} {\bibinfo {author} {\bibfnamefont {N.}~\bibnamefont
  {Chamel}}\ and\ \bibinfo {author} {\bibfnamefont {P.}~\bibnamefont
  {Haensel}},\ }\href {\doibase 10.12942/lrr-2008-10} {\bibfield  {journal}
  {\bibinfo  {journal} {Living Rev. Rel.}\ }\textbf {\bibinfo {volume} {11}},\
  \bibinfo {pages} {10} (\bibinfo {year} {2008})},\ \Eprint
  {http://arxiv.org/abs/0812.3955} {arXiv:0812.3955 [astro-ph]} \BibitemShut
  {NoStop}%
\bibitem [{\citenamefont {Israel}\ \emph {et~al.}(2005)\citenamefont {Israel},
  \citenamefont {Belloni}, \citenamefont {Stella}, \citenamefont {Rephaeli},
  \citenamefont {Gruber} \emph {et~al.}}]{Israel:2005}%
  \BibitemOpen
  \bibfield  {author} {\bibinfo {author} {\bibfnamefont {G.}~\bibnamefont
  {Israel}}, \bibinfo {author} {\bibfnamefont {T.}~\bibnamefont {Belloni}},
  \bibinfo {author} {\bibfnamefont {L.}~\bibnamefont {Stella}}, \bibinfo
  {author} {\bibfnamefont {Y.}~\bibnamefont {Rephaeli}}, \bibinfo {author}
  {\bibfnamefont {D.}~\bibnamefont {Gruber}},  \emph {et~al.},\ }\href
  {\doibase 10.1086/432615} {\bibfield  {journal} {\bibinfo  {journal}
  {Astrophys. J.}\ }\textbf {\bibinfo {volume} {628}},\ \bibinfo {pages} {L53}
  (\bibinfo {year} {2005})},\ \Eprint {http://arxiv.org/abs/astro-ph/0505255}
  {arXiv:astro-ph/0505255 [astro-ph]} \BibitemShut {NoStop}%
\bibitem [{\citenamefont {Strohmayer}\ and\ \citenamefont
  {Watts}(2005)}]{Strohmayer:2005}%
  \BibitemOpen
  \bibfield  {author} {\bibinfo {author} {\bibfnamefont {T.~E.}\ \bibnamefont
  {Strohmayer}}\ and\ \bibinfo {author} {\bibfnamefont {A.~L.}\ \bibnamefont
  {Watts}},\ }\href {\doibase 10.1086/497911} {\bibfield  {journal} {\bibinfo
  {journal} {Astrophys. J.}\ }\textbf {\bibinfo {volume} {632}},\ \bibinfo
  {pages} {L111} (\bibinfo {year} {2005})},\ \Eprint
  {http://arxiv.org/abs/astro-ph/0508206} {arXiv:astro-ph/0508206 [astro-ph]}
  \BibitemShut {NoStop}%
\bibitem [{\citenamefont {Watts}\ and\ \citenamefont
  {Strohmayer}(2006)}]{Watts:2006}%
  \BibitemOpen
  \bibfield  {author} {\bibinfo {author} {\bibfnamefont {A.~L.}\ \bibnamefont
  {Watts}}\ and\ \bibinfo {author} {\bibfnamefont {T.~E.}\ \bibnamefont
  {Strohmayer}},\ }\href {\doibase 10.1086/500735} {\bibfield  {journal}
  {\bibinfo  {journal} {Astrophys. J.}\ }\textbf {\bibinfo {volume} {637}},\
  \bibinfo {pages} {L117} (\bibinfo {year} {2006})},\ \Eprint
  {http://arxiv.org/abs/astro-ph/0512630} {arXiv:astro-ph/0512630 [astro-ph]}
  \BibitemShut {NoStop}%
\bibitem [{\citenamefont {Strohmayer}\ and\ \citenamefont
  {Watts}(2006)}]{Strohmayer:2006}%
  \BibitemOpen
  \bibfield  {author} {\bibinfo {author} {\bibfnamefont {T.~E.}\ \bibnamefont
  {Strohmayer}}\ and\ \bibinfo {author} {\bibfnamefont {A.~L.}\ \bibnamefont
  {Watts}},\ }\href {\doibase 10.1086/508703} {\bibfield  {journal} {\bibinfo
  {journal} {Astrophys. J.}\ }\textbf {\bibinfo {volume} {653}},\ \bibinfo
  {pages} {593} (\bibinfo {year} {2006})},\ \Eprint
  {http://arxiv.org/abs/astro-ph/0608463} {arXiv:astro-ph/0608463 [astro-ph]}
  \BibitemShut {NoStop}%
\bibitem [{\citenamefont {Hambaryan}\ \emph {et~al.}(2011)\citenamefont
  {Hambaryan}, \citenamefont {Neuhaeuser},\ and\ \citenamefont
  {Kokkotas}}]{Hambaryan:2010}%
  \BibitemOpen
  \bibfield  {author} {\bibinfo {author} {\bibfnamefont {V.}~\bibnamefont
  {Hambaryan}}, \bibinfo {author} {\bibfnamefont {R.}~\bibnamefont
  {Neuhaeuser}}, \ and\ \bibinfo {author} {\bibfnamefont {K.~D.}\ \bibnamefont
  {Kokkotas}},\ }\href {\doibase 10.1051/0004-6361/201015273} {\bibfield
  {journal} {\bibinfo  {journal} {Astron. Astrophys.}\ }\textbf {\bibinfo
  {volume} {528}},\ \bibinfo {pages} {A45} (\bibinfo {year} {2011})},\ \Eprint
  {http://arxiv.org/abs/1012.5654} {arXiv:1012.5654 [astro-ph.SR]} \BibitemShut
  {NoStop}%
\bibitem [{\citenamefont {Duncan}(1998)}]{Duncan:1998my}%
  \BibitemOpen
  \bibfield  {author} {\bibinfo {author} {\bibfnamefont {R.~C.}\ \bibnamefont
  {Duncan}},\ }\href {\doibase 10.1086/311303} {\bibfield  {journal} {\bibinfo
  {journal} {Astrophys. J.}\ }\textbf {\bibinfo {volume} {498}},\ \bibinfo
  {pages} {L45} (\bibinfo {year} {1998})},\ \Eprint
  {http://arxiv.org/abs/astro-ph/9803060} {arXiv:astro-ph/9803060 [astro-ph]}
  \BibitemShut {NoStop}%
\bibitem [{\citenamefont {Levin}(2006)}]{Levin:2006ck}%
  \BibitemOpen
  \bibfield  {author} {\bibinfo {author} {\bibfnamefont {Y.}~\bibnamefont
  {Levin}},\ }\href {\doibase 10.1111/j.1745-3933.2006.00155.x} {\bibfield
  {journal} {\bibinfo  {journal} {Mon. Not. Roy. Astron. Soc.}\ }\textbf
  {\bibinfo {volume} {368}},\ \bibinfo {pages} {L35} (\bibinfo {year}
  {2006})},\ \Eprint {http://arxiv.org/abs/astro-ph/0601020}
  {arXiv:astro-ph/0601020 [astro-ph]} \BibitemShut {NoStop}%
\bibitem [{\citenamefont {Gabler}\ \emph {et~al.}(2011)\citenamefont {Gabler},
  \citenamefont {Cerda-Duran}, \citenamefont {Font}, \citenamefont {Muller},\
  and\ \citenamefont {Stergioulas}}]{Gabler:2010rp}%
  \BibitemOpen
  \bibfield  {author} {\bibinfo {author} {\bibfnamefont {M.}~\bibnamefont
  {Gabler}}, \bibinfo {author} {\bibfnamefont {P.}~\bibnamefont {Cerda-Duran}},
  \bibinfo {author} {\bibfnamefont {J.~A.}\ \bibnamefont {Font}}, \bibinfo
  {author} {\bibfnamefont {E.}~\bibnamefont {Muller}}, \ and\ \bibinfo {author}
  {\bibfnamefont {N.}~\bibnamefont {Stergioulas}},\ }\href {\doibase
  10.1111/j.1745-3933.2010.00974.x} {\bibfield  {journal} {\bibinfo  {journal}
  {Mon. Not. Roy. Astron. Soc.}\ }\textbf {\bibinfo {volume} {410}},\ \bibinfo
  {pages} {37} (\bibinfo {year} {2011})},\ \Eprint
  {http://arxiv.org/abs/1007.0856} {arXiv:1007.0856 [astro-ph.HE]} \BibitemShut
  {NoStop}%
\bibitem [{\citenamefont {Andersson}\ \emph {et~al.}(2009)\citenamefont
  {Andersson}, \citenamefont {Glampedakis},\ and\ \citenamefont
  {Samuelsson}}]{Andersson:2008}%
  \BibitemOpen
  \bibfield  {author} {\bibinfo {author} {\bibfnamefont {N.}~\bibnamefont
  {Andersson}}, \bibinfo {author} {\bibfnamefont {K.}~\bibnamefont
  {Glampedakis}}, \ and\ \bibinfo {author} {\bibfnamefont {L.}~\bibnamefont
  {Samuelsson}},\ }\href {\doibase 10.1111/j.1365-2966.2009.14734.x} {\bibfield
   {journal} {\bibinfo  {journal} {Mon. Not. Roy. Astron. Soc.}\ }\textbf
  {\bibinfo {volume} {396}},\ \bibinfo {pages} {894} (\bibinfo {year}
  {2009})},\ \Eprint {http://arxiv.org/abs/0812.2417} {arXiv:0812.2417
  [astro-ph]} \BibitemShut {NoStop}%
\bibitem [{\citenamefont {Samuelsson}\ and\ \citenamefont
  {Andersson}(2009)}]{Samuelsson:2009xz}%
  \BibitemOpen
  \bibfield  {author} {\bibinfo {author} {\bibfnamefont {L.}~\bibnamefont
  {Samuelsson}}\ and\ \bibinfo {author} {\bibfnamefont {N.}~\bibnamefont
  {Andersson}},\ }\href {\doibase 10.1088/0264-9381/26/15/155016} {\bibfield
  {journal} {\bibinfo  {journal} {Class. Quant. Grav.}\ }\textbf {\bibinfo
  {volume} {26}},\ \bibinfo {pages} {155016} (\bibinfo {year} {2009})},\
  \Eprint {http://arxiv.org/abs/0903.2437} {arXiv:0903.2437 [astro-ph.SR]}
  \BibitemShut {NoStop}%
\bibitem [{\citenamefont {Passamonti}\ and\ \citenamefont
  {Andersson}(2012)}]{Passamonti:2011mc}%
  \BibitemOpen
  \bibfield  {author} {\bibinfo {author} {\bibfnamefont {A.}~\bibnamefont
  {Passamonti}}\ and\ \bibinfo {author} {\bibfnamefont {N.}~\bibnamefont
  {Andersson}},\ }\href {\doibase 10.1111/j.1365-2966.2011.19725.x} {\bibfield
  {journal} {\bibinfo  {journal} {Mon. Not. Roy. Astron. Soc.}\ }\textbf
  {\bibinfo {volume} {419}},\ \bibinfo {pages} {638} (\bibinfo {year}
  {2012})},\ \Eprint {http://arxiv.org/abs/1105.4787} {arXiv:1105.4787
  [astro-ph.SR]} \BibitemShut {NoStop}%
\bibitem [{\citenamefont {Sotani}\ \emph {et~al.}(2013)\citenamefont {Sotani},
  \citenamefont {Nakazato}, \citenamefont {Iida},\ and\ \citenamefont
  {Oyamatsu}}]{Sotani:2013jya}%
  \BibitemOpen
  \bibfield  {author} {\bibinfo {author} {\bibfnamefont {H.}~\bibnamefont
  {Sotani}}, \bibinfo {author} {\bibfnamefont {K.}~\bibnamefont {Nakazato}},
  \bibinfo {author} {\bibfnamefont {K.}~\bibnamefont {Iida}}, \ and\ \bibinfo
  {author} {\bibfnamefont {K.}~\bibnamefont {Oyamatsu}},\ }\href {\doibase
  10.1093/mnras/stt1152} {\bibfield  {journal} {\bibinfo  {journal} {Mon. Not.
  Roy. Astron. Soc.}\ }\textbf {\bibinfo {volume} {434}},\ \bibinfo {pages}
  {2060} (\bibinfo {year} {2013})},\ \Eprint {http://arxiv.org/abs/1303.4500}
  {arXiv:1303.4500 [astro-ph.HE]} \BibitemShut {NoStop}%
\bibitem [{\citenamefont {Chamel}(2012)}]{Chamel:2012zn}%
  \BibitemOpen
  \bibfield  {author} {\bibinfo {author} {\bibfnamefont {N.}~\bibnamefont
  {Chamel}},\ }\href {\doibase 10.1103/PhysRevC.85.035801,
  10.1103/PhysRevC.85.039902} {\bibfield  {journal} {\bibinfo  {journal} {Phys.
  Rev.}\ }\textbf {\bibinfo {volume} {C85}},\ \bibinfo {pages} {035801}
  (\bibinfo {year} {2012})},\ \Eprint {http://arxiv.org/abs/1203.0119}
  {arXiv:1203.0119 [nucl-th]} \BibitemShut {NoStop}%
\bibitem [{\citenamefont {Gearheart}\ \emph {et~al.}(2011)\citenamefont
  {Gearheart}, \citenamefont {Newton}, \citenamefont {Hooker},\ and\
  \citenamefont {Li}}]{Gearheart:2011qt}%
  \BibitemOpen
  \bibfield  {author} {\bibinfo {author} {\bibfnamefont {M.}~\bibnamefont
  {Gearheart}}, \bibinfo {author} {\bibfnamefont {W.~G.}\ \bibnamefont
  {Newton}}, \bibinfo {author} {\bibfnamefont {J.}~\bibnamefont {Hooker}}, \
  and\ \bibinfo {author} {\bibfnamefont {B.-A.}\ \bibnamefont {Li}},\ }\href
  {\doibase 10.1111/j.1365-2966.2011.19628.x} {\bibfield  {journal} {\bibinfo
  {journal} {Mon. Not. Roy. Astron. Soc.}\ }\textbf {\bibinfo {volume} {418}},\
  \bibinfo {pages} {2343} (\bibinfo {year} {2011})},\ \Eprint
  {http://arxiv.org/abs/1106.4875} {arXiv:1106.4875 [astro-ph.SR]} \BibitemShut
  {NoStop}%
\bibitem [{\citenamefont {Sotani}(2011)}]{Sotani:2011nn}%
  \BibitemOpen
  \bibfield  {author} {\bibinfo {author} {\bibfnamefont {H.}~\bibnamefont
  {Sotani}},\ }\href {\doibase 10.1111/j.1745-3933.2011.01122.x} {\bibfield
  {journal} {\bibinfo  {journal} {Mon. Not. Roy. Astron. Soc.}\ }\textbf
  {\bibinfo {volume} {417}},\ \bibinfo {pages} {L70} (\bibinfo {year}
  {2011})},\ \Eprint {http://arxiv.org/abs/1106.2621} {arXiv:1106.2621
  [astro-ph.HE]} \BibitemShut {NoStop}%
\bibitem [{\citenamefont {Passamonti}\ and\ \citenamefont
  {Pons}(2016)}]{Passamonti:2016jfo}%
  \BibitemOpen
  \bibfield  {author} {\bibinfo {author} {\bibfnamefont {A.}~\bibnamefont
  {Passamonti}}\ and\ \bibinfo {author} {\bibfnamefont {J.~A.}\ \bibnamefont
  {Pons}},\ }\href {\doibase 10.1093/mnras/stw1880} {\bibfield  {journal}
  {\bibinfo  {journal} {Mon. Not. Roy. Astron. Soc.}\ }\textbf {\bibinfo
  {volume} {463}},\ \bibinfo {pages} {1173} (\bibinfo {year} {2016})},\ \Eprint
  {http://arxiv.org/abs/1606.02132} {arXiv:1606.02132 [astro-ph.HE]}
  \BibitemShut {NoStop}%
\bibitem [{\citenamefont {Steiner}\ and\ \citenamefont
  {Watts}(2009)}]{SteinerWatts2009}%
  \BibitemOpen
  \bibfield  {author} {\bibinfo {author} {\bibfnamefont {A.~W.}\ \bibnamefont
  {Steiner}}\ and\ \bibinfo {author} {\bibfnamefont {A.~L.}\ \bibnamefont
  {Watts}},\ }\href {\doibase 10.1103/PhysRevLett.103.181101} {\bibfield
  {journal} {\bibinfo  {journal} {Phys. Rev. Lett.}\ }\textbf {\bibinfo
  {volume} {103}},\ \bibinfo {pages} {181101} (\bibinfo {year} {2009})},\
  \Eprint {http://arxiv.org/abs/0902.1683} {arXiv:0902.1683 [astro-ph.HE]}
  \BibitemShut {NoStop}%
\bibitem [{\citenamefont {Baym}\ \emph {et~al.}(1971)\citenamefont {Baym},
  \citenamefont {Pethick},\ and\ \citenamefont {Sutherland}}]{BPS}%
  \BibitemOpen
  \bibfield  {author} {\bibinfo {author} {\bibfnamefont {G.}~\bibnamefont
  {Baym}}, \bibinfo {author} {\bibfnamefont {C.}~\bibnamefont {Pethick}}, \
  and\ \bibinfo {author} {\bibfnamefont {P.}~\bibnamefont {Sutherland}},\
  }\href {\doibase 10.1086/151216} {\bibfield  {journal} {\bibinfo  {journal}
  {Astrophys. J.}\ }\textbf {\bibinfo {volume} {170}},\ \bibinfo {pages} {299}
  (\bibinfo {year} {1971})}\BibitemShut {NoStop}%
\bibitem [{\citenamefont {Ruester}\ \emph {et~al.}(2006)\citenamefont
  {Ruester}, \citenamefont {Hempel},\ and\ \citenamefont
  {Schaffner-Bielich}}]{Ruester:2005}%
  \BibitemOpen
  \bibfield  {author} {\bibinfo {author} {\bibfnamefont {S.~B.}\ \bibnamefont
  {Ruester}}, \bibinfo {author} {\bibfnamefont {M.}~\bibnamefont {Hempel}}, \
  and\ \bibinfo {author} {\bibfnamefont {J.}~\bibnamefont
  {Schaffner-Bielich}},\ }\href {\doibase 10.1103/PhysRevC.73.035804}
  {\bibfield  {journal} {\bibinfo  {journal} {Phys. Rev.}\ }\textbf {\bibinfo
  {volume} {C73}},\ \bibinfo {pages} {035804} (\bibinfo {year} {2006})},\
  \Eprint {http://arxiv.org/abs/astro-ph/0509325} {arXiv:astro-ph/0509325
  [astro-ph]} \BibitemShut {NoStop}%
\bibitem [{\citenamefont {Epelbaum}\ \emph {et~al.}(2009)\citenamefont
  {Epelbaum}, \citenamefont {Hammer},\ and\ \citenamefont
  {Mei{\ss}ner}}]{Epelbaum:2009a}%
  \BibitemOpen
  \bibfield  {author} {\bibinfo {author} {\bibfnamefont {E.}~\bibnamefont
  {Epelbaum}}, \bibinfo {author} {\bibfnamefont {H.-W.}\ \bibnamefont
  {Hammer}}, \ and\ \bibinfo {author} {\bibfnamefont {U.-G.}\ \bibnamefont
  {Mei{\ss}ner}},\ }\href {\doibase 10.1103/RevModPhys.81.1773} {\bibfield
  {journal} {\bibinfo  {journal} {Rev. Mod. Phys.}\ }\textbf {\bibinfo {volume}
  {81}},\ \bibinfo {pages} {1773} (\bibinfo {year} {2009})},\ \Eprint
  {http://arxiv.org/abs/0811.1338} {arXiv:0811.1338 [nucl-th]} \BibitemShut
  {NoStop}%
\bibitem [{\citenamefont {Machleidt}\ and\ \citenamefont
  {Entem}(2011)}]{Entem:2011}%
  \BibitemOpen
  \bibfield  {author} {\bibinfo {author} {\bibfnamefont {R.}~\bibnamefont
  {Machleidt}}\ and\ \bibinfo {author} {\bibfnamefont {D.~R.}\ \bibnamefont
  {Entem}},\ }\href {\doibase 10.1016/j.physrep.2011.02.001} {\bibfield
  {journal} {\bibinfo  {journal} {Phys. Rept.}\ }\textbf {\bibinfo {volume}
  {503}},\ \bibinfo {pages} {1} (\bibinfo {year} {2011})},\ \Eprint
  {http://arxiv.org/abs/1105.2919} {arXiv:1105.2919 [nucl-th]} \BibitemShut
  {NoStop}%
\bibitem [{\citenamefont {Hebeler}\ \emph {et~al.}(2015)\citenamefont
  {Hebeler}, \citenamefont {Holt}, \citenamefont {Menendez},\ and\
  \citenamefont {Schwenk}}]{Hebeler:review2015}%
  \BibitemOpen
  \bibfield  {author} {\bibinfo {author} {\bibfnamefont {K.}~\bibnamefont
  {Hebeler}}, \bibinfo {author} {\bibfnamefont {J.~D.}\ \bibnamefont {Holt}},
  \bibinfo {author} {\bibfnamefont {J.}~\bibnamefont {Menendez}}, \ and\
  \bibinfo {author} {\bibfnamefont {A.}~\bibnamefont {Schwenk}},\ }\href
  {\doibase 10.1146/annurev-nucl-102313-025446} {\bibfield  {journal} {\bibinfo
   {journal} {Ann. Rev. Nucl. Part. Sci.}\ }\textbf {\bibinfo {volume} {65}},\
  \bibinfo {pages} {457} (\bibinfo {year} {2015})},\ \Eprint
  {http://arxiv.org/abs/1508.06893} {arXiv:1508.06893 [nucl-th]} \BibitemShut
  {NoStop}%
\bibitem [{\citenamefont {Hebeler}\ and\ \citenamefont
  {Schwenk}(2010)}]{Hebeler:2010a}%
  \BibitemOpen
  \bibfield  {author} {\bibinfo {author} {\bibfnamefont {K.}~\bibnamefont
  {Hebeler}}\ and\ \bibinfo {author} {\bibfnamefont {A.}~\bibnamefont
  {Schwenk}},\ }\href {\doibase 10.1103/PhysRevC.82.014314} {\bibfield
  {journal} {\bibinfo  {journal} {Phys. Rev.}\ }\textbf {\bibinfo {volume}
  {C82}},\ \bibinfo {pages} {014314} (\bibinfo {year} {2010})},\ \Eprint
  {http://arxiv.org/abs/0911.0483} {arXiv:0911.0483 [nucl-th]} \BibitemShut
  {NoStop}%
\bibitem [{\citenamefont {Hebeler}\ \emph {et~al.}(2011)\citenamefont
  {Hebeler}, \citenamefont {Bogner}, \citenamefont {Furnstahl}, \citenamefont
  {Nogga},\ and\ \citenamefont {Schwenk}}]{nucmatt}%
  \BibitemOpen
  \bibfield  {author} {\bibinfo {author} {\bibfnamefont {K.}~\bibnamefont
  {Hebeler}}, \bibinfo {author} {\bibfnamefont {S.~K.}\ \bibnamefont {Bogner}},
  \bibinfo {author} {\bibfnamefont {R.~J.}\ \bibnamefont {Furnstahl}}, \bibinfo
  {author} {\bibfnamefont {A.}~\bibnamefont {Nogga}}, \ and\ \bibinfo {author}
  {\bibfnamefont {A.}~\bibnamefont {Schwenk}},\ }\href {\doibase
  10.1103/PhysRevC.83.031301} {\bibfield  {journal} {\bibinfo  {journal} {Phys.
  Rev.}\ }\textbf {\bibinfo {volume} {C83}},\ \bibinfo {pages} {031301}
  (\bibinfo {year} {2011})},\ \Eprint {http://arxiv.org/abs/1012.3381}
  {arXiv:1012.3381 [nucl-th]} \BibitemShut {NoStop}%
\bibitem [{\citenamefont {Holt}\ \emph
  {et~al.}(2013{\natexlab{a}})\citenamefont {Holt}, \citenamefont {Kaiser},\
  and\ \citenamefont {Weise}}]{Holt2:2012}%
  \BibitemOpen
  \bibfield  {author} {\bibinfo {author} {\bibfnamefont {J.~W.}\ \bibnamefont
  {Holt}}, \bibinfo {author} {\bibfnamefont {N.}~\bibnamefont {Kaiser}}, \ and\
  \bibinfo {author} {\bibfnamefont {W.}~\bibnamefont {Weise}},\ }\href
  {\doibase 10.1103/PhysRevC.87.014338} {\bibfield  {journal} {\bibinfo
  {journal} {Phys. Rev.}\ }\textbf {\bibinfo {volume} {C87}},\ \bibinfo {pages}
  {014338} (\bibinfo {year} {2013}{\natexlab{a}})},\ \Eprint
  {http://arxiv.org/abs/1209.5296} {arXiv:1209.5296 [nucl-th]} \BibitemShut
  {NoStop}%
\bibitem [{\citenamefont {Tews}\ \emph {et~al.}(2013)\citenamefont {Tews},
  \citenamefont {Kr\"uger}, \citenamefont {Hebeler},\ and\ \citenamefont
  {Schwenk}}]{Tews:2013}%
  \BibitemOpen
  \bibfield  {author} {\bibinfo {author} {\bibfnamefont {I.}~\bibnamefont
  {Tews}}, \bibinfo {author} {\bibfnamefont {T.}~\bibnamefont {Kr\"uger}},
  \bibinfo {author} {\bibfnamefont {K.}~\bibnamefont {Hebeler}}, \ and\
  \bibinfo {author} {\bibfnamefont {A.}~\bibnamefont {Schwenk}},\ }\href
  {\doibase 10.1103/PhysRevLett.110.032504} {\bibfield  {journal} {\bibinfo
  {journal} {Phys. Rev. Lett.}\ }\textbf {\bibinfo {volume} {110}},\ \bibinfo
  {pages} {032504} (\bibinfo {year} {2013})},\ \Eprint
  {http://arxiv.org/abs/1206.0025} {arXiv:1206.0025 [nucl-th]} \BibitemShut
  {NoStop}%
\bibitem [{\citenamefont {Baardsen}\ \emph {et~al.}(2013)\citenamefont
  {Baardsen}, \citenamefont {Ekstr\"om}, \citenamefont {Hagen},\ and\
  \citenamefont {Hjorth-Jensen}}]{CCnucmatt}%
  \BibitemOpen
  \bibfield  {author} {\bibinfo {author} {\bibfnamefont {G.}~\bibnamefont
  {Baardsen}}, \bibinfo {author} {\bibfnamefont {A.}~\bibnamefont {Ekstr\"om}},
  \bibinfo {author} {\bibfnamefont {G.}~\bibnamefont {Hagen}}, \ and\ \bibinfo
  {author} {\bibfnamefont {M.}~\bibnamefont {Hjorth-Jensen}},\ }\href {\doibase
  10.1103/PhysRevC.88.054312} {\bibfield  {journal} {\bibinfo  {journal} {Phys.
  Rev.}\ }\textbf {\bibinfo {volume} {C88}},\ \bibinfo {pages} {054312}
  (\bibinfo {year} {2013})},\ \Eprint {http://arxiv.org/abs/1306.5681}
  {arXiv:1306.5681 [nucl-th]} \BibitemShut {NoStop}%
\bibitem [{\citenamefont {Drischler}\ \emph {et~al.}(2014)\citenamefont
  {Drischler}, \citenamefont {Som\`{a}},\ and\ \citenamefont
  {Schwenk}}]{Drischler:2013}%
  \BibitemOpen
  \bibfield  {author} {\bibinfo {author} {\bibfnamefont {C.}~\bibnamefont
  {Drischler}}, \bibinfo {author} {\bibfnamefont {V.}~\bibnamefont {Som\`{a}}},
  \ and\ \bibinfo {author} {\bibfnamefont {A.}~\bibnamefont {Schwenk}},\ }\href
  {\doibase 10.1103/PhysRevC.89.025806} {\bibfield  {journal} {\bibinfo
  {journal} {Phys. Rev.}\ }\textbf {\bibinfo {volume} {C89}},\ \bibinfo {pages}
  {025806} (\bibinfo {year} {2014})},\ \Eprint {http://arxiv.org/abs/1310.5627}
  {arXiv:1310.5627 [nucl-th]} \BibitemShut {NoStop}%
\bibitem [{\citenamefont {Hagen}\ \emph {et~al.}(2014)\citenamefont {Hagen},
  \citenamefont {Papenbrock}, \citenamefont {Ekstr\"om}, \citenamefont {Wendt},
  \citenamefont {Baardsen} \emph {et~al.}}]{Hagen:2014}%
  \BibitemOpen
  \bibfield  {author} {\bibinfo {author} {\bibfnamefont {G.}~\bibnamefont
  {Hagen}}, \bibinfo {author} {\bibfnamefont {T.}~\bibnamefont {Papenbrock}},
  \bibinfo {author} {\bibfnamefont {A.}~\bibnamefont {Ekstr\"om}}, \bibinfo
  {author} {\bibfnamefont {K.~A.}\ \bibnamefont {Wendt}}, \bibinfo {author}
  {\bibfnamefont {G.}~\bibnamefont {Baardsen}},  \emph {et~al.},\ }\href
  {\doibase 10.1103/PhysRevC.89.014319} {\bibfield  {journal} {\bibinfo
  {journal} {Phys. Rev.}\ }\textbf {\bibinfo {volume} {C89}},\ \bibinfo {pages}
  {014319} (\bibinfo {year} {2014})},\ \Eprint {http://arxiv.org/abs/1311.2925}
  {arXiv:1311.2925 [nucl-th]} \BibitemShut {NoStop}%
\bibitem [{\citenamefont {Roggero}\ \emph {et~al.}(2014)\citenamefont
  {Roggero}, \citenamefont {Mukherjee},\ and\ \citenamefont
  {Pederiva}}]{Roggero:2014}%
  \BibitemOpen
  \bibfield  {author} {\bibinfo {author} {\bibfnamefont {A.}~\bibnamefont
  {Roggero}}, \bibinfo {author} {\bibfnamefont {A.}~\bibnamefont {Mukherjee}},
  \ and\ \bibinfo {author} {\bibfnamefont {F.}~\bibnamefont {Pederiva}},\
  }\href {\doibase 10.1103/PhysRevLett.112.221103} {\bibfield  {journal}
  {\bibinfo  {journal} {Phys. Rev. Lett.}\ }\textbf {\bibinfo {volume} {112}},\
  \bibinfo {pages} {221103} (\bibinfo {year} {2014})},\ \Eprint
  {http://arxiv.org/abs/1402.1576} {arXiv:1402.1576 [nucl-th]} \BibitemShut
  {NoStop}%
\bibitem [{\citenamefont {Wlazłowski}\ \emph {et~al.}(2014)\citenamefont
  {Wlazłowski}, \citenamefont {Holt}, \citenamefont {Moroz}, \citenamefont
  {Bulgac},\ and\ \citenamefont {Roche}}]{Wlazlowski:2014jna}%
  \BibitemOpen
  \bibfield  {author} {\bibinfo {author} {\bibfnamefont {G.}~\bibnamefont
  {Wlazłowski}}, \bibinfo {author} {\bibfnamefont {J.~W.}\ \bibnamefont
  {Holt}}, \bibinfo {author} {\bibfnamefont {S.}~\bibnamefont {Moroz}},
  \bibinfo {author} {\bibfnamefont {A.}~\bibnamefont {Bulgac}}, \ and\ \bibinfo
  {author} {\bibfnamefont {K.~J.}\ \bibnamefont {Roche}},\ }\href {\doibase
  10.1103/PhysRevLett.113.182503} {\bibfield  {journal} {\bibinfo  {journal}
  {Phys. Rev. Lett.}\ }\textbf {\bibinfo {volume} {113}},\ \bibinfo {pages}
  {182503} (\bibinfo {year} {2014})},\ \Eprint {http://arxiv.org/abs/1403.3753}
  {arXiv:1403.3753 [nucl-th]} \BibitemShut {NoStop}%
\bibitem [{\citenamefont {Lynn}\ \emph {et~al.}(2014)\citenamefont {Lynn},
  \citenamefont {Carlson}, \citenamefont {Epelbaum}, \citenamefont {Gandolfi},
  \citenamefont {Gezerlis} \emph {et~al.}}]{Lynn2014}%
  \BibitemOpen
  \bibfield  {author} {\bibinfo {author} {\bibfnamefont {J.~E.}\ \bibnamefont
  {Lynn}}, \bibinfo {author} {\bibfnamefont {J.}~\bibnamefont {Carlson}},
  \bibinfo {author} {\bibfnamefont {E.}~\bibnamefont {Epelbaum}}, \bibinfo
  {author} {\bibfnamefont {S.}~\bibnamefont {Gandolfi}}, \bibinfo {author}
  {\bibfnamefont {A.}~\bibnamefont {Gezerlis}},  \emph {et~al.},\ }\href
  {\doibase 10.1103/PhysRevLett.113.192501} {\bibfield  {journal} {\bibinfo
  {journal} {Phys. Rev. Lett.}\ }\textbf {\bibinfo {volume} {113}},\ \bibinfo
  {pages} {192501} (\bibinfo {year} {2014})},\ \Eprint
  {http://arxiv.org/abs/1406.2787} {arXiv:1406.2787 [nucl-th]} \BibitemShut
  {NoStop}%
\bibitem [{\citenamefont {Carbone}\ \emph {et~al.}(2014)\citenamefont
  {Carbone}, \citenamefont {Rios},\ and\ \citenamefont {Polls}}]{Carbone:2014}%
  \BibitemOpen
  \bibfield  {author} {\bibinfo {author} {\bibfnamefont {A.}~\bibnamefont
  {Carbone}}, \bibinfo {author} {\bibfnamefont {A.}~\bibnamefont {Rios}}, \
  and\ \bibinfo {author} {\bibfnamefont {A.}~\bibnamefont {Polls}},\ }\href
  {\doibase 10.1103/PhysRevC.90.054322} {\bibfield  {journal} {\bibinfo
  {journal} {Phys. Rev.}\ }\textbf {\bibinfo {volume} {C90}},\ \bibinfo {pages}
  {054322} (\bibinfo {year} {2014})},\ \Eprint {http://arxiv.org/abs/1408.0717}
  {arXiv:1408.0717 [nucl-th]} \BibitemShut {NoStop}%
\bibitem [{\citenamefont {Lynn}\ \emph {et~al.}(2016)\citenamefont {Lynn},
  \citenamefont {Tews}, \citenamefont {Carlson}, \citenamefont {Gandolfi},
  \citenamefont {Gezerlis}, \citenamefont {Schmidt},\ and\ \citenamefont
  {Schwenk}}]{Lynn:2015}%
  \BibitemOpen
  \bibfield  {author} {\bibinfo {author} {\bibfnamefont {J.~E.}\ \bibnamefont
  {Lynn}}, \bibinfo {author} {\bibfnamefont {I.}~\bibnamefont {Tews}}, \bibinfo
  {author} {\bibfnamefont {J.}~\bibnamefont {Carlson}}, \bibinfo {author}
  {\bibfnamefont {S.}~\bibnamefont {Gandolfi}}, \bibinfo {author}
  {\bibfnamefont {A.}~\bibnamefont {Gezerlis}}, \bibinfo {author}
  {\bibfnamefont {K.~E.}\ \bibnamefont {Schmidt}}, \ and\ \bibinfo {author}
  {\bibfnamefont {A.}~\bibnamefont {Schwenk}},\ }\href {\doibase
  10.1103/PhysRevLett.116.062501} {\bibfield  {journal} {\bibinfo  {journal}
  {Phys. Rev. Lett.}\ }\textbf {\bibinfo {volume} {116}},\ \bibinfo {pages}
  {062501} (\bibinfo {year} {2016})},\ \Eprint
  {http://arxiv.org/abs/1509.03470} {arXiv:1509.03470 [nucl-th]} \BibitemShut
  {NoStop}%
\bibitem [{\citenamefont {Drischler}\ \emph {et~al.}(2016)\citenamefont
  {Drischler}, \citenamefont {Hebeler},\ and\ \citenamefont
  {Schwenk}}]{Drischler:2015}%
  \BibitemOpen
  \bibfield  {author} {\bibinfo {author} {\bibfnamefont {C.}~\bibnamefont
  {Drischler}}, \bibinfo {author} {\bibfnamefont {K.}~\bibnamefont {Hebeler}},
  \ and\ \bibinfo {author} {\bibfnamefont {A.}~\bibnamefont {Schwenk}},\ }\href
  {\doibase 10.1103/PhysRevC.93.054314} {\bibfield  {journal} {\bibinfo
  {journal} {Phys. Rev.}\ }\textbf {\bibinfo {volume} {C93}},\ \bibinfo {pages}
  {054314} (\bibinfo {year} {2016})},\ \Eprint
  {http://arxiv.org/abs/1510.06728} {arXiv:1510.06728 [nucl-th]} \BibitemShut
  {NoStop}%
\bibitem [{\citenamefont {Otsuka}\ \emph {et~al.}(2010)\citenamefont {Otsuka},
  \citenamefont {Suzuki}, \citenamefont {Holt}, \citenamefont {Schwenk},\ and\
  \citenamefont {Akaishi}}]{SM}%
  \BibitemOpen
  \bibfield  {author} {\bibinfo {author} {\bibfnamefont {T.}~\bibnamefont
  {Otsuka}}, \bibinfo {author} {\bibfnamefont {T.}~\bibnamefont {Suzuki}},
  \bibinfo {author} {\bibfnamefont {J.~D.}\ \bibnamefont {Holt}}, \bibinfo
  {author} {\bibfnamefont {A.}~\bibnamefont {Schwenk}}, \ and\ \bibinfo
  {author} {\bibfnamefont {Y.}~\bibnamefont {Akaishi}},\ }\href {\doibase
  10.1103/PhysRevLett.105.032501} {\bibfield  {journal} {\bibinfo  {journal}
  {Phys. Rev. Lett.}\ }\textbf {\bibinfo {volume} {105}},\ \bibinfo {pages}
  {032501} (\bibinfo {year} {2010})},\ \Eprint {http://arxiv.org/abs/0908.2607}
  {arXiv:0908.2607 [nucl-th]} \BibitemShut {NoStop}%
\bibitem [{\citenamefont {Hagen}\ \emph
  {et~al.}(2012{\natexlab{a}})\citenamefont {Hagen}, \citenamefont
  {Hjorth-Jensen}, \citenamefont {Jansen}, \citenamefont {Machleidt},\ and\
  \citenamefont {Papenbrock}}]{CC1}%
  \BibitemOpen
  \bibfield  {author} {\bibinfo {author} {\bibfnamefont {G.}~\bibnamefont
  {Hagen}}, \bibinfo {author} {\bibfnamefont {M.}~\bibnamefont
  {Hjorth-Jensen}}, \bibinfo {author} {\bibfnamefont {G.~R.}\ \bibnamefont
  {Jansen}}, \bibinfo {author} {\bibfnamefont {R.}~\bibnamefont {Machleidt}}, \
  and\ \bibinfo {author} {\bibfnamefont {T.}~\bibnamefont {Papenbrock}},\
  }\href {\doibase 10.1103/PhysRevLett.108.242501} {\bibfield  {journal}
  {\bibinfo  {journal} {Phys. Rev. Lett.}\ }\textbf {\bibinfo {volume} {108}},\
  \bibinfo {pages} {242501} (\bibinfo {year} {2012}{\natexlab{a}})},\ \Eprint
  {http://arxiv.org/abs/1202.2839} {arXiv:1202.2839 [nucl-th]} \BibitemShut
  {NoStop}%
\bibitem [{\citenamefont {Hagen}\ \emph
  {et~al.}(2012{\natexlab{b}})\citenamefont {Hagen}, \citenamefont
  {Hjorth-Jensen}, \citenamefont {Jansen}, \citenamefont {Machleidt},\ and\
  \citenamefont {Papenbrock}}]{CC1b}%
  \BibitemOpen
  \bibfield  {author} {\bibinfo {author} {\bibfnamefont {G.}~\bibnamefont
  {Hagen}}, \bibinfo {author} {\bibfnamefont {M.}~\bibnamefont
  {Hjorth-Jensen}}, \bibinfo {author} {\bibfnamefont {G.~R.}\ \bibnamefont
  {Jansen}}, \bibinfo {author} {\bibfnamefont {R.}~\bibnamefont {Machleidt}}, \
  and\ \bibinfo {author} {\bibfnamefont {T.}~\bibnamefont {Papenbrock}},\
  }\href {\doibase 10.1103/PhysRevLett.109.032502} {\bibfield  {journal}
  {\bibinfo  {journal} {Phys. Rev. Lett.}\ }\textbf {\bibinfo {volume} {109}},\
  \bibinfo {pages} {032502} (\bibinfo {year} {2012}{\natexlab{b}})},\ \Eprint
  {http://arxiv.org/abs/1204.3612} {arXiv:1204.3612 [nucl-th]} \BibitemShut
  {NoStop}%
\bibitem [{\citenamefont {Cipollone}\ \emph {et~al.}(2013)\citenamefont
  {Cipollone}, \citenamefont {Barbieri},\ and\ \citenamefont
  {Navrátil}}]{Cipollone:2013zma}%
  \BibitemOpen
  \bibfield  {author} {\bibinfo {author} {\bibfnamefont {A.}~\bibnamefont
  {Cipollone}}, \bibinfo {author} {\bibfnamefont {C.}~\bibnamefont {Barbieri}},
  \ and\ \bibinfo {author} {\bibfnamefont {P.}~\bibnamefont {Navrátil}},\
  }\href {\doibase 10.1103/PhysRevLett.111.062501} {\bibfield  {journal}
  {\bibinfo  {journal} {Phys. Rev. Lett.}\ }\textbf {\bibinfo {volume} {111}},\
  \bibinfo {pages} {062501} (\bibinfo {year} {2013})},\ \Eprint
  {http://arxiv.org/abs/1303.4900} {arXiv:1303.4900 [nucl-th]} \BibitemShut
  {NoStop}%
\bibitem [{\citenamefont {Hergert}\ \emph {et~al.}(2013)\citenamefont
  {Hergert}, \citenamefont {Binder}, \citenamefont {Calci}, \citenamefont
  {Langhammer},\ and\ \citenamefont {Roth}}]{Hergert:2013b}%
  \BibitemOpen
  \bibfield  {author} {\bibinfo {author} {\bibfnamefont {H.}~\bibnamefont
  {Hergert}}, \bibinfo {author} {\bibfnamefont {S.}~\bibnamefont {Binder}},
  \bibinfo {author} {\bibfnamefont {A.}~\bibnamefont {Calci}}, \bibinfo
  {author} {\bibfnamefont {J.}~\bibnamefont {Langhammer}}, \ and\ \bibinfo
  {author} {\bibfnamefont {R.}~\bibnamefont {Roth}},\ }\href {\doibase
  10.1103/PhysRevLett.110.242501} {\bibfield  {journal} {\bibinfo  {journal}
  {Phys. Rev. Lett.}\ }\textbf {\bibinfo {volume} {110}},\ \bibinfo {pages}
  {242501} (\bibinfo {year} {2013})},\ \Eprint {http://arxiv.org/abs/1302.7294}
  {arXiv:1302.7294 [nucl-th]} \BibitemShut {NoStop}%
\bibitem [{\citenamefont {Holt}\ \emph
  {et~al.}(2013{\natexlab{b}})\citenamefont {Holt}, \citenamefont {Menendez},\
  and\ \citenamefont {Schwenk}}]{Holt:2011fj}%
  \BibitemOpen
  \bibfield  {author} {\bibinfo {author} {\bibfnamefont {J.~D.}\ \bibnamefont
  {Holt}}, \bibinfo {author} {\bibfnamefont {J.}~\bibnamefont {Menendez}}, \
  and\ \bibinfo {author} {\bibfnamefont {A.}~\bibnamefont {Schwenk}},\ }\href
  {\doibase 10.1140/epja/i2013-13039-2} {\bibfield  {journal} {\bibinfo
  {journal} {Eur. Phys. J.}\ }\textbf {\bibinfo {volume} {A49}},\ \bibinfo
  {pages} {39} (\bibinfo {year} {2013}{\natexlab{b}})},\ \Eprint
  {http://arxiv.org/abs/1108.2680} {arXiv:1108.2680 [nucl-th]} \BibitemShut
  {NoStop}%
\bibitem [{\citenamefont {Holt}\ \emph {et~al.}(2012)\citenamefont {Holt},
  \citenamefont {Otsuka}, \citenamefont {Schwenk},\ and\ \citenamefont
  {Suzuki}}]{Holt:2010yb}%
  \BibitemOpen
  \bibfield  {author} {\bibinfo {author} {\bibfnamefont {J.~D.}\ \bibnamefont
  {Holt}}, \bibinfo {author} {\bibfnamefont {T.}~\bibnamefont {Otsuka}},
  \bibinfo {author} {\bibfnamefont {A.}~\bibnamefont {Schwenk}}, \ and\
  \bibinfo {author} {\bibfnamefont {T.}~\bibnamefont {Suzuki}},\ }\href
  {\doibase 10.1088/0954-3899/39/8/085111} {\bibfield  {journal} {\bibinfo
  {journal} {J. Phys.}\ }\textbf {\bibinfo {volume} {G39}},\ \bibinfo {pages}
  {085111} (\bibinfo {year} {2012})},\ \Eprint {http://arxiv.org/abs/1009.5984}
  {arXiv:1009.5984 [nucl-th]} \BibitemShut {NoStop}%
\bibitem [{\citenamefont {Wienholtz}\ \emph {et~al.}(2013)\citenamefont
  {Wienholtz}, \citenamefont {Beck}, \citenamefont {Blaum}, \citenamefont
  {Borgmann}, \citenamefont {Breitenfeldt} \emph {et~al.}}]{Ca}%
  \BibitemOpen
  \bibfield  {author} {\bibinfo {author} {\bibfnamefont {F.}~\bibnamefont
  {Wienholtz}}, \bibinfo {author} {\bibfnamefont {D.}~\bibnamefont {Beck}},
  \bibinfo {author} {\bibfnamefont {K.}~\bibnamefont {Blaum}}, \bibinfo
  {author} {\bibfnamefont {C.}~\bibnamefont {Borgmann}}, \bibinfo {author}
  {\bibfnamefont {M.}~\bibnamefont {Breitenfeldt}},  \emph {et~al.},\ }\href
  {\doibase 10.1038/nature12226} {\bibfield  {journal} {\bibinfo  {journal}
  {Nature}\ }\textbf {\bibinfo {volume} {498}},\ \bibinfo {pages} {346}
  (\bibinfo {year} {2013})}\BibitemShut {NoStop}%
\bibitem [{\citenamefont {Hergert}\ \emph {et~al.}(2014)\citenamefont
  {Hergert}, \citenamefont {Bogner}, \citenamefont {Morris}, \citenamefont
  {Binder}, \citenamefont {Calci}, \citenamefont {Langhammer},\ and\
  \citenamefont {Roth}}]{Hergert:2014iaa}%
  \BibitemOpen
  \bibfield  {author} {\bibinfo {author} {\bibfnamefont {H.}~\bibnamefont
  {Hergert}}, \bibinfo {author} {\bibfnamefont {S.~K.}\ \bibnamefont {Bogner}},
  \bibinfo {author} {\bibfnamefont {T.~D.}\ \bibnamefont {Morris}}, \bibinfo
  {author} {\bibfnamefont {S.}~\bibnamefont {Binder}}, \bibinfo {author}
  {\bibfnamefont {A.}~\bibnamefont {Calci}}, \bibinfo {author} {\bibfnamefont
  {J.}~\bibnamefont {Langhammer}}, \ and\ \bibinfo {author} {\bibfnamefont
  {R.}~\bibnamefont {Roth}},\ }\href {\doibase 10.1103/PhysRevC.90.041302}
  {\bibfield  {journal} {\bibinfo  {journal} {Phys. Rev.}\ }\textbf {\bibinfo
  {volume} {C90}},\ \bibinfo {pages} {041302} (\bibinfo {year} {2014})},\
  \Eprint {http://arxiv.org/abs/1408.6555} {arXiv:1408.6555 [nucl-th]}
  \BibitemShut {NoStop}%
\bibitem [{\citenamefont {Som\`{a}}\ \emph {et~al.}(2014)\citenamefont
  {Som\`{a}}, \citenamefont {Cipollone}, \citenamefont {Barbieri},
  \citenamefont {Navrátil},\ and\ \citenamefont {Duguet}}]{Soma:2013xha}%
  \BibitemOpen
  \bibfield  {author} {\bibinfo {author} {\bibfnamefont {V.}~\bibnamefont
  {Som\`{a}}}, \bibinfo {author} {\bibfnamefont {A.}~\bibnamefont {Cipollone}},
  \bibinfo {author} {\bibfnamefont {C.}~\bibnamefont {Barbieri}}, \bibinfo
  {author} {\bibfnamefont {P.}~\bibnamefont {Navrátil}}, \ and\ \bibinfo
  {author} {\bibfnamefont {T.}~\bibnamefont {Duguet}},\ }\href {\doibase
  10.1103/PhysRevC.89.061301} {\bibfield  {journal} {\bibinfo  {journal} {Phys.
  Rev.}\ }\textbf {\bibinfo {volume} {C89}},\ \bibinfo {pages} {061301}
  (\bibinfo {year} {2014})},\ \Eprint {http://arxiv.org/abs/1312.2068}
  {arXiv:1312.2068 [nucl-th]} \BibitemShut {NoStop}%
\bibitem [{\citenamefont {Epelbaum}\ \emph {et~al.}(2014)\citenamefont
  {Epelbaum}, \citenamefont {Krebs}, \citenamefont {L\"ahde}, \citenamefont
  {Lee}, \citenamefont {Mei{\ss}ner} \emph {et~al.}}]{nuclattice2}%
  \BibitemOpen
  \bibfield  {author} {\bibinfo {author} {\bibfnamefont {E.}~\bibnamefont
  {Epelbaum}}, \bibinfo {author} {\bibfnamefont {H.}~\bibnamefont {Krebs}},
  \bibinfo {author} {\bibfnamefont {T.~A.}\ \bibnamefont {L\"ahde}}, \bibinfo
  {author} {\bibfnamefont {D.}~\bibnamefont {Lee}}, \bibinfo {author}
  {\bibfnamefont {U.-G.}\ \bibnamefont {Mei{\ss}ner}},  \emph {et~al.},\ }\href
  {\doibase 10.1103/PhysRevLett.112.102501} {\bibfield  {journal} {\bibinfo
  {journal} {Phys. Rev. Lett.}\ }\textbf {\bibinfo {volume} {112}},\ \bibinfo
  {pages} {102501} (\bibinfo {year} {2014})},\ \Eprint
  {http://arxiv.org/abs/1312.7703} {arXiv:1312.7703 [nucl-th]} \BibitemShut
  {NoStop}%
\bibitem [{\citenamefont {Hagen}\ \emph {et~al.}(2015)\citenamefont {Hagen}
  \emph {et~al.}}]{Hagen:2015yea}%
  \BibitemOpen
  \bibfield  {author} {\bibinfo {author} {\bibfnamefont {G.}~\bibnamefont
  {Hagen}} \emph {et~al.},\ }\href {\doibase 10.1038/nphys3529} {\bibfield
  {journal} {\bibinfo  {journal} {Nature Phys.}\ }\textbf {\bibinfo {volume}
  {12}},\ \bibinfo {pages} {186} (\bibinfo {year} {2015})},\ \Eprint
  {http://arxiv.org/abs/1509.07169} {arXiv:1509.07169 [nucl-th]} \BibitemShut
  {NoStop}%
\bibitem [{\citenamefont {Garcia~Ruiz}\ \emph {et~al.}(2016)\citenamefont
  {Garcia~Ruiz} \emph {et~al.}}]{Ruiz:2016gne}%
  \BibitemOpen
  \bibfield  {author} {\bibinfo {author} {\bibfnamefont {R.~F.}\ \bibnamefont
  {Garcia~Ruiz}} \emph {et~al.},\ }\href {\doibase 10.1038/nphys3645}
  {\bibfield  {journal} {\bibinfo  {journal} {Nature Phys.}\ }\textbf {\bibinfo
  {volume} {12}},\ \bibinfo {pages} {594} (\bibinfo {year} {2016})},\ \Eprint
  {http://arxiv.org/abs/1602.07906} {arXiv:1602.07906 [nucl-ex]} \BibitemShut
  {NoStop}%
\bibitem [{\citenamefont {Kr\"uger}\ \emph {et~al.}(2013)\citenamefont
  {Kr\"uger}, \citenamefont {Tews}, \citenamefont {Hebeler},\ and\
  \citenamefont {Schwenk}}]{Kruger:2013kua}%
  \BibitemOpen
  \bibfield  {author} {\bibinfo {author} {\bibfnamefont {T.}~\bibnamefont
  {Kr\"uger}}, \bibinfo {author} {\bibfnamefont {I.}~\bibnamefont {Tews}},
  \bibinfo {author} {\bibfnamefont {K.}~\bibnamefont {Hebeler}}, \ and\
  \bibinfo {author} {\bibfnamefont {A.}~\bibnamefont {Schwenk}},\ }\href
  {\doibase 10.1103/PhysRevC.88.025802} {\bibfield  {journal} {\bibinfo
  {journal} {Phys. Rev.}\ }\textbf {\bibinfo {volume} {C88}},\ \bibinfo {pages}
  {025802} (\bibinfo {year} {2013})},\ \Eprint {http://arxiv.org/abs/1304.2212}
  {arXiv:1304.2212 [nucl-th]} \BibitemShut {NoStop}%
\bibitem [{\citenamefont {Gandolfi}\ \emph {et~al.}(2012)\citenamefont
  {Gandolfi}, \citenamefont {Carlson},\ and\ \citenamefont
  {Reddy}}]{Gandolfi:2012}%
  \BibitemOpen
  \bibfield  {author} {\bibinfo {author} {\bibfnamefont {S.}~\bibnamefont
  {Gandolfi}}, \bibinfo {author} {\bibfnamefont {J.}~\bibnamefont {Carlson}}, \
  and\ \bibinfo {author} {\bibfnamefont {S.}~\bibnamefont {Reddy}},\ }\href
  {\doibase 10.1103/PhysRevC.85.032801} {\bibfield  {journal} {\bibinfo
  {journal} {Phys. Rev.}\ }\textbf {\bibinfo {volume} {C85}},\ \bibinfo {pages}
  {032801} (\bibinfo {year} {2012})},\ \Eprint {http://arxiv.org/abs/1101.1921}
  {arXiv:1101.1921 [nucl-th]} \BibitemShut {NoStop}%
\bibitem [{\citenamefont {Lattimer}\ and\ \citenamefont
  {Prakash}(2016)}]{Lattimer:2015nhk}%
  \BibitemOpen
  \bibfield  {author} {\bibinfo {author} {\bibfnamefont {J.~M.}\ \bibnamefont
  {Lattimer}}\ and\ \bibinfo {author} {\bibfnamefont {M.}~\bibnamefont
  {Prakash}},\ }\href {\doibase 10.1016/j.physrep.2015.12.005} {\bibfield
  {journal} {\bibinfo  {journal} {Phys. Rept.}\ }\textbf {\bibinfo {volume}
  {621}},\ \bibinfo {pages} {127} (\bibinfo {year} {2016})},\ \Eprint
  {http://arxiv.org/abs/1512.07820} {arXiv:1512.07820 [astro-ph.SR]}
  \BibitemShut {NoStop}%
\bibitem [{\citenamefont {Lattimer}\ and\ \citenamefont
  {Lim}(2013)}]{Lattimer2012}%
  \BibitemOpen
  \bibfield  {author} {\bibinfo {author} {\bibfnamefont {J.~M.}\ \bibnamefont
  {Lattimer}}\ and\ \bibinfo {author} {\bibfnamefont {Y.}~\bibnamefont {Lim}},\
  }\href {\doibase 10.1088/0004-637X/771/1/51} {\bibfield  {journal} {\bibinfo
  {journal} {Astrophys. J.}\ }\textbf {\bibinfo {volume} {771}},\ \bibinfo
  {pages} {51} (\bibinfo {year} {2013})},\ \Eprint
  {http://arxiv.org/abs/1203.4286} {arXiv:1203.4286 [nucl-th]} \BibitemShut
  {NoStop}%
\bibitem [{\citenamefont {Steiner}(2008)}]{Steiner:2007}%
  \BibitemOpen
  \bibfield  {author} {\bibinfo {author} {\bibfnamefont {A.~W.}\ \bibnamefont
  {Steiner}},\ }\href {\doibase 10.1103/PhysRevC.77.035805} {\bibfield
  {journal} {\bibinfo  {journal} {Phys. Rev.}\ }\textbf {\bibinfo {volume}
  {C77}},\ \bibinfo {pages} {035805} (\bibinfo {year} {2008})},\ \Eprint
  {http://arxiv.org/abs/0711.1812} {arXiv:0711.1812 [nucl-th]} \BibitemShut
  {NoStop}%
\bibitem [{\citenamefont {Audi}\ \emph {et~al.}(2014)\citenamefont {Audi},
  \citenamefont {Wang}, \citenamefont {Wapstra}, \citenamefont {Kondev},
  \citenamefont {MacCormick},\ and\ \citenamefont {Xu}}]{Audi:2014}%
  \BibitemOpen
  \bibfield  {author} {\bibinfo {author} {\bibfnamefont {G.}~\bibnamefont
  {Audi}}, \bibinfo {author} {\bibfnamefont {M.}~\bibnamefont {Wang}}, \bibinfo
  {author} {\bibfnamefont {A.~H.}\ \bibnamefont {Wapstra}}, \bibinfo {author}
  {\bibfnamefont {F.~G.}\ \bibnamefont {Kondev}}, \bibinfo {author}
  {\bibfnamefont {M.}~\bibnamefont {MacCormick}}, \ and\ \bibinfo {author}
  {\bibfnamefont {X.}~\bibnamefont {Xu}},\ }\href {\doibase
  10.1016/j.nds.2014.06.126} {\bibfield  {journal} {\bibinfo  {journal} {Nucl.
  Data Sheets}\ }\textbf {\bibinfo {volume} {120}},\ \bibinfo {pages} {1}
  (\bibinfo {year} {2014})}\BibitemShut {NoStop}%
\bibitem [{\citenamefont {Douchin}\ and\ \citenamefont
  {Haensel}(2001)}]{Douchin:2001sv}%
  \BibitemOpen
  \bibfield  {author} {\bibinfo {author} {\bibfnamefont {F.}~\bibnamefont
  {Douchin}}\ and\ \bibinfo {author} {\bibfnamefont {P.}~\bibnamefont
  {Haensel}},\ }\href {\doibase 10.1051/0004-6361:20011402} {\bibfield
  {journal} {\bibinfo  {journal} {Astron. Astrophys.}\ }\textbf {\bibinfo
  {volume} {380}},\ \bibinfo {pages} {151} (\bibinfo {year} {2001})},\ \Eprint
  {http://arxiv.org/abs/astro-ph/0111092} {arXiv:astro-ph/0111092 [astro-ph]}
  \BibitemShut {NoStop}%
\bibitem [{\citenamefont {Negele}\ and\ \citenamefont
  {Vautherin}(1973)}]{Negele:1971vb}%
  \BibitemOpen
  \bibfield  {author} {\bibinfo {author} {\bibfnamefont {J.~W.}\ \bibnamefont
  {Negele}}\ and\ \bibinfo {author} {\bibfnamefont {D.}~\bibnamefont
  {Vautherin}},\ }\href {\doibase 10.1016/0375-9474(73)90349-7} {\bibfield
  {journal} {\bibinfo  {journal} {Nucl. Phys.}\ }\textbf {\bibinfo {volume}
  {A207}},\ \bibinfo {pages} {298} (\bibinfo {year} {1973})}\BibitemShut
  {NoStop}%
\bibitem [{\citenamefont {Deibel}\ \emph {et~al.}(2014)\citenamefont {Deibel},
  \citenamefont {Steiner},\ and\ \citenamefont {Brown}}]{Deibel:2013sia}%
  \BibitemOpen
  \bibfield  {author} {\bibinfo {author} {\bibfnamefont {A.~T.}\ \bibnamefont
  {Deibel}}, \bibinfo {author} {\bibfnamefont {A.~W.}\ \bibnamefont {Steiner}},
  \ and\ \bibinfo {author} {\bibfnamefont {E.~F.}\ \bibnamefont {Brown}},\
  }\href {\doibase 10.1103/PhysRevC.90.025802} {\bibfield  {journal} {\bibinfo
  {journal} {Phys. Rev.}\ }\textbf {\bibinfo {volume} {C90}},\ \bibinfo {pages}
  {025802} (\bibinfo {year} {2014})},\ \Eprint {http://arxiv.org/abs/1303.3270}
  {arXiv:1303.3270 [astro-ph.HE]} \BibitemShut {NoStop}%
\bibitem [{\citenamefont {Ravenhall}\ \emph {et~al.}(1983)\citenamefont
  {Ravenhall}, \citenamefont {Pethick},\ and\ \citenamefont
  {Wilson}}]{Ravenhall:1983}%
  \BibitemOpen
  \bibfield  {author} {\bibinfo {author} {\bibfnamefont {D.~G.}\ \bibnamefont
  {Ravenhall}}, \bibinfo {author} {\bibfnamefont {C.~J.}\ \bibnamefont
  {Pethick}}, \ and\ \bibinfo {author} {\bibfnamefont {J.~R.}\ \bibnamefont
  {Wilson}},\ }\href {\doibase 10.1103/PhysRevLett.50.2066} {\bibfield
  {journal} {\bibinfo  {journal} {Phys. Rev. Lett.}\ }\textbf {\bibinfo
  {volume} {50}},\ \bibinfo {pages} {2066} (\bibinfo {year}
  {1983})}\BibitemShut {NoStop}%
\bibitem [{\citenamefont {Strohmayer}\ \emph {et~al.}(1991)\citenamefont
  {Strohmayer}, \citenamefont {van Horn}, \citenamefont {Ogata}, \citenamefont
  {Iyetomi},\ and\ \citenamefont {Ichimaru}}]{Strohmayer:1991}%
  \BibitemOpen
  \bibfield  {author} {\bibinfo {author} {\bibfnamefont {T.}~\bibnamefont
  {Strohmayer}}, \bibinfo {author} {\bibfnamefont {H.~M.}\ \bibnamefont {van
  Horn}}, \bibinfo {author} {\bibfnamefont {S.}~\bibnamefont {Ogata}}, \bibinfo
  {author} {\bibfnamefont {H.}~\bibnamefont {Iyetomi}}, \ and\ \bibinfo
  {author} {\bibfnamefont {S.}~\bibnamefont {Ichimaru}},\ }\href@noop {}
  {\bibfield  {journal} {\bibinfo  {journal} {ApJ.}\ }\textbf {\bibinfo
  {volume} {375}},\ \bibinfo {pages} {679} (\bibinfo {year}
  {1991})}\BibitemShut {NoStop}%
\bibitem [{\citenamefont {Farouki}\ and\ \citenamefont
  {Hamaguchi}(1993)}]{Farouki:1993}%
  \BibitemOpen
  \bibfield  {author} {\bibinfo {author} {\bibfnamefont {R.~T.}\ \bibnamefont
  {Farouki}}\ and\ \bibinfo {author} {\bibfnamefont {S.}~\bibnamefont
  {Hamaguchi}},\ }\href {\doibase 10.1103/PhysRevE.47.4330} {\bibfield
  {journal} {\bibinfo  {journal} {Phys. Rev.}\ }\textbf {\bibinfo {volume}
  {E47}},\ \bibinfo {pages} {4330} (\bibinfo {year} {1993})}\BibitemShut
  {NoStop}%
\bibitem [{\citenamefont {Kobyakov}\ and\ \citenamefont
  {Pethick}(2013)}]{Kobyakov:2013}%
  \BibitemOpen
  \bibfield  {author} {\bibinfo {author} {\bibfnamefont {D.}~\bibnamefont
  {Kobyakov}}\ and\ \bibinfo {author} {\bibfnamefont {C.~J.}\ \bibnamefont
  {Pethick}},\ }\href {\doibase 10.1103/PhysRevC.87.055803} {\bibfield
  {journal} {\bibinfo  {journal} {Phys. Rev.}\ }\textbf {\bibinfo {volume}
  {C87}},\ \bibinfo {pages} {055803} (\bibinfo {year} {2013})},\ \Eprint
  {http://arxiv.org/abs/1303.1315} {arXiv:1303.1315 [nucl-th]} \BibitemShut
  {NoStop}%
\bibitem [{\citenamefont {Lander}\ \emph {et~al.}(2015)\citenamefont {Lander},
  \citenamefont {Andersson}, \citenamefont {Antonopoulou},\ and\ \citenamefont
  {Watts}}]{Lander:2014}%
  \BibitemOpen
  \bibfield  {author} {\bibinfo {author} {\bibfnamefont {S.~K.}\ \bibnamefont
  {Lander}}, \bibinfo {author} {\bibfnamefont {N.}~\bibnamefont {Andersson}},
  \bibinfo {author} {\bibfnamefont {D.}~\bibnamefont {Antonopoulou}}, \ and\
  \bibinfo {author} {\bibfnamefont {A.~L.}\ \bibnamefont {Watts}},\ }\href
  {\doibase 10.1093/mnras/stv432} {\bibfield  {journal} {\bibinfo  {journal}
  {Mon. Not. Roy. Astron. Soc.}\ }\textbf {\bibinfo {volume} {449}},\ \bibinfo
  {pages} {2047} (\bibinfo {year} {2015})},\ \Eprint
  {http://arxiv.org/abs/1412.5852} {arXiv:1412.5852 [astro-ph.HE]} \BibitemShut
  {NoStop}%
\bibitem [{\citenamefont {Hebeler}\ \emph {et~al.}(2010)\citenamefont
  {Hebeler}, \citenamefont {Lattimer}, \citenamefont {Pethick},\ and\
  \citenamefont {Schwenk}}]{Hebeler:2010b}%
  \BibitemOpen
  \bibfield  {author} {\bibinfo {author} {\bibfnamefont {K.}~\bibnamefont
  {Hebeler}}, \bibinfo {author} {\bibfnamefont {J.~M.}\ \bibnamefont
  {Lattimer}}, \bibinfo {author} {\bibfnamefont {C.~J.}\ \bibnamefont
  {Pethick}}, \ and\ \bibinfo {author} {\bibfnamefont {A.}~\bibnamefont
  {Schwenk}},\ }\href {\doibase 10.1103/PhysRevLett.105.161102} {\bibfield
  {journal} {\bibinfo  {journal} {Phys. Rev. Lett.}\ }\textbf {\bibinfo
  {volume} {105}},\ \bibinfo {pages} {161102} (\bibinfo {year} {2010})},\
  \Eprint {http://arxiv.org/abs/1007.1746} {arXiv:1007.1746 [nucl-th]}
  \BibitemShut {NoStop}%
\bibitem [{\citenamefont {Piro}(2005)}]{Piro:2005}%
  \BibitemOpen
  \bibfield  {author} {\bibinfo {author} {\bibfnamefont {A.~L.}\ \bibnamefont
  {Piro}},\ }\href {\doibase 10.1086/499049} {\bibfield  {journal} {\bibinfo
  {journal} {Astrophys. J.}\ }\textbf {\bibinfo {volume} {634}},\ \bibinfo
  {pages} {L153} (\bibinfo {year} {2005})},\ \Eprint
  {http://arxiv.org/abs/astro-ph/0510578} {arXiv:astro-ph/0510578 [astro-ph]}
  \BibitemShut {NoStop}%
\bibitem [{\citenamefont {Samuelsson}\ and\ \citenamefont
  {Andersson}(2007)}]{Samuelsson:2006tt}%
  \BibitemOpen
  \bibfield  {author} {\bibinfo {author} {\bibfnamefont {L.}~\bibnamefont
  {Samuelsson}}\ and\ \bibinfo {author} {\bibfnamefont {N.}~\bibnamefont
  {Andersson}},\ }\href {\doibase 10.1111/j.1365-2966.2006.11147.x} {\bibfield
  {journal} {\bibinfo  {journal} {Mon. Not. Roy. Astron. Soc.}\ }\textbf
  {\bibinfo {volume} {374}},\ \bibinfo {pages} {256} (\bibinfo {year}
  {2007})},\ \Eprint {http://arxiv.org/abs/astro-ph/0609265}
  {arXiv:astro-ph/0609265 [astro-ph]} \BibitemShut {NoStop}%
\bibitem [{\citenamefont {Glampedakis}\ \emph {et~al.}(2006)\citenamefont
  {Glampedakis}, \citenamefont {Samuelsson},\ and\ \citenamefont
  {Andersson}}]{Glampedakis:2006}%
  \BibitemOpen
  \bibfield  {author} {\bibinfo {author} {\bibfnamefont {K.}~\bibnamefont
  {Glampedakis}}, \bibinfo {author} {\bibfnamefont {L.}~\bibnamefont
  {Samuelsson}}, \ and\ \bibinfo {author} {\bibfnamefont {N.}~\bibnamefont
  {Andersson}},\ }\href {\doibase 10.1111/j.1745-3933.2006.00211.x} {\bibfield
  {journal} {\bibinfo  {journal} {Mon. Not. Roy. Astron. Soc.}\ }\textbf
  {\bibinfo {volume} {371}},\ \bibinfo {pages} {L74} (\bibinfo {year}
  {2006})},\ \Eprint {http://arxiv.org/abs/astro-ph/0605461}
  {arXiv:astro-ph/0605461 [astro-ph]} \BibitemShut {NoStop}%
\bibitem [{\citenamefont {Levin}(2007)}]{Levin:2006qd}%
  \BibitemOpen
  \bibfield  {author} {\bibinfo {author} {\bibfnamefont {Y.}~\bibnamefont
  {Levin}},\ }\href {\doibase 10.1111/j.1365-2966.2007.11582.x} {\bibfield
  {journal} {\bibinfo  {journal} {Mon. Not. Roy. Astron. Soc.}\ }\textbf
  {\bibinfo {volume} {377}},\ \bibinfo {pages} {159} (\bibinfo {year}
  {2007})},\ \Eprint {http://arxiv.org/abs/astro-ph/0612725}
  {arXiv:astro-ph/0612725 [astro-ph]} \BibitemShut {NoStop}%
\bibitem [{\citenamefont {Colaiuda}\ and\ \citenamefont
  {Kokkotas}(2011)}]{Colaiuda:2010pc}%
  \BibitemOpen
  \bibfield  {author} {\bibinfo {author} {\bibfnamefont {A.}~\bibnamefont
  {Colaiuda}}\ and\ \bibinfo {author} {\bibfnamefont {K.~D.}\ \bibnamefont
  {Kokkotas}},\ }\href {\doibase 10.1111/j.1365-2966.2011.18602.x} {\bibfield
  {journal} {\bibinfo  {journal} {Mon. Not. Roy. Astron. Soc.}\ }\textbf
  {\bibinfo {volume} {414}},\ \bibinfo {pages} {3014} (\bibinfo {year}
  {2011})},\ \Eprint {http://arxiv.org/abs/1012.3103} {arXiv:1012.3103 [gr-qc]}
  \BibitemShut {NoStop}%
\bibitem [{\citenamefont {van Hoven}\ and\ \citenamefont
  {Levin}(2011)}]{vanHoven:2010gy}%
  \BibitemOpen
  \bibfield  {author} {\bibinfo {author} {\bibfnamefont {M.}~\bibnamefont {van
  Hoven}}\ and\ \bibinfo {author} {\bibfnamefont {Y.}~\bibnamefont {Levin}},\
  }\href {\doibase 10.1111/j.1365-2966.2010.17499.x} {\bibfield  {journal}
  {\bibinfo  {journal} {Mon. Not. Roy. Astron. Soc.}\ }\textbf {\bibinfo
  {volume} {410}},\ \bibinfo {pages} {1036} (\bibinfo {year} {2011})},\ \Eprint
  {http://arxiv.org/abs/1006.0348} {arXiv:1006.0348 [astro-ph.HE]} \BibitemShut
  {NoStop}%
\bibitem [{\citenamefont {Gabler}\ \emph {et~al.}(2012)\citenamefont {Gabler},
  \citenamefont {Duran}, \citenamefont {Stergioulas}, \citenamefont {Font},\
  and\ \citenamefont {Muller}}]{Gabler:2011am}%
  \BibitemOpen
  \bibfield  {author} {\bibinfo {author} {\bibfnamefont {M.}~\bibnamefont
  {Gabler}}, \bibinfo {author} {\bibfnamefont {P.~C.}\ \bibnamefont {Duran}},
  \bibinfo {author} {\bibfnamefont {N.}~\bibnamefont {Stergioulas}}, \bibinfo
  {author} {\bibfnamefont {J.~A.}\ \bibnamefont {Font}}, \ and\ \bibinfo
  {author} {\bibfnamefont {E.}~\bibnamefont {Muller}},\ }\href {\doibase
  10.1111/j.1365-2966.2012.20454.x} {\bibfield  {journal} {\bibinfo  {journal}
  {Mon. Not. Roy. Astron. Soc.}\ }\textbf {\bibinfo {volume} {421}},\ \bibinfo
  {pages} {2054} (\bibinfo {year} {2012})},\ \Eprint
  {http://arxiv.org/abs/1109.6233} {arXiv:1109.6233 [astro-ph.HE]} \BibitemShut
  {NoStop}%
\bibitem [{\citenamefont {Gabler}\ \emph {et~al.}(2013)\citenamefont {Gabler},
  \citenamefont {Cerda-Duran}, \citenamefont {Font}, \citenamefont {Muller},\
  and\ \citenamefont {Stergioulas}}]{Gabler:2012jh}%
  \BibitemOpen
  \bibfield  {author} {\bibinfo {author} {\bibfnamefont {M.}~\bibnamefont
  {Gabler}}, \bibinfo {author} {\bibfnamefont {P.}~\bibnamefont {Cerda-Duran}},
  \bibinfo {author} {\bibfnamefont {J.~A.}\ \bibnamefont {Font}}, \bibinfo
  {author} {\bibfnamefont {E.}~\bibnamefont {Muller}}, \ and\ \bibinfo {author}
  {\bibfnamefont {N.}~\bibnamefont {Stergioulas}},\ }\href {\doibase
  10.1093/mnras/sts721} {\bibfield  {journal} {\bibinfo  {journal} {Mon. Not.
  Roy. Astron. Soc.}\ }\textbf {\bibinfo {volume} {430}},\ \bibinfo {pages}
  {1811} (\bibinfo {year} {2013})},\ \Eprint {http://arxiv.org/abs/1208.6443}
  {arXiv:1208.6443 [astro-ph.SR]} \BibitemShut {NoStop}%
\bibitem [{\citenamefont {Gabler}\ \emph {et~al.}(2016)\citenamefont {Gabler},
  \citenamefont {Cerdá-Durán}, \citenamefont {Stergioulas}, \citenamefont
  {Font},\ and\ \citenamefont {Müller}}]{Gabler:2016rth}%
  \BibitemOpen
  \bibfield  {author} {\bibinfo {author} {\bibfnamefont {M.}~\bibnamefont
  {Gabler}}, \bibinfo {author} {\bibfnamefont {P.}~\bibnamefont
  {Cerdá-Durán}}, \bibinfo {author} {\bibfnamefont {N.}~\bibnamefont
  {Stergioulas}}, \bibinfo {author} {\bibfnamefont {J.~A.}\ \bibnamefont
  {Font}}, \ and\ \bibinfo {author} {\bibfnamefont {E.}~\bibnamefont
  {Müller}},\ }\href {\doibase 10.1093/mnras/stw1272} {\bibfield  {journal}
  {\bibinfo  {journal} {Mon. Not. Roy. Astron. Soc.}\ }\textbf {\bibinfo
  {volume} {460}},\ \bibinfo {pages} {4242} (\bibinfo {year} {2016})},\ \Eprint
  {http://arxiv.org/abs/1605.07638} {arXiv:1605.07638 [astro-ph.HE]}
  \BibitemShut {NoStop}%
\bibitem [{\citenamefont {Steiner}\ \emph {et~al.}(2015)\citenamefont
  {Steiner}, \citenamefont {Gandolfi}, \citenamefont {Fattoyev},\ and\
  \citenamefont {Newton}}]{Steiner:2014pda}%
  \BibitemOpen
  \bibfield  {author} {\bibinfo {author} {\bibfnamefont {A.~W.}\ \bibnamefont
  {Steiner}}, \bibinfo {author} {\bibfnamefont {S.}~\bibnamefont {Gandolfi}},
  \bibinfo {author} {\bibfnamefont {F.~J.}\ \bibnamefont {Fattoyev}}, \ and\
  \bibinfo {author} {\bibfnamefont {W.~G.}\ \bibnamefont {Newton}},\ }\href
  {\doibase 10.1103/PhysRevC.91.015804} {\bibfield  {journal} {\bibinfo
  {journal} {Phys. Rev.}\ }\textbf {\bibinfo {volume} {C91}},\ \bibinfo {pages}
  {015804} (\bibinfo {year} {2015})},\ \Eprint {http://arxiv.org/abs/1403.7546}
  {arXiv:1403.7546 [nucl-th]} \BibitemShut {NoStop}%
\bibitem [{\citenamefont {Sotani}\ \emph {et~al.}(2007)\citenamefont {Sotani},
  \citenamefont {Kokkotas},\ and\ \citenamefont {Stergioulas}}]{Sotani:2006at}%
  \BibitemOpen
  \bibfield  {author} {\bibinfo {author} {\bibfnamefont {H.}~\bibnamefont
  {Sotani}}, \bibinfo {author} {\bibfnamefont {K.~D.}\ \bibnamefont
  {Kokkotas}}, \ and\ \bibinfo {author} {\bibfnamefont {N.}~\bibnamefont
  {Stergioulas}},\ }\href {\doibase 10.1111/j.1365-2966.2006.11304.x}
  {\bibfield  {journal} {\bibinfo  {journal} {Mon. Not. Roy. Astron. Soc.}\
  }\textbf {\bibinfo {volume} {375}},\ \bibinfo {pages} {261} (\bibinfo {year}
  {2007})},\ \Eprint {http://arxiv.org/abs/astro-ph/0608626}
  {arXiv:astro-ph/0608626 [astro-ph]} \BibitemShut {NoStop}%
\end{thebibliography}%
\end{document}